\documentclass[12pt,a4paper]{article}
\pdfoutput=1
\usepackage{graphicx}
\usepackage[T1]{fontenc}
\usepackage[sc,osf]{mathpazo}
\usepackage{a4wide}  
\usepackage{latexsym,amsthm,amsfonts,amsmath,mathrsfs,amssymb}
\usepackage{booktabs} 
\usepackage[unicode,implicit]{hyperref}
\hypersetup{%
  pdftitle    = {$\alpha'$-corrected non-Abelian black holes in string theory}
  pdfkeywords = {Yang-Mills, instanton, black hole, string theory, entropy,
    branes, supergravity, supersymmetry, corrections, higher-order},
  pdfauthor   = {Pablo A. Cano, Patrick Meessen, Tomas Ortin, Pedro F. Ramirez},
  plainpages  = true,
  colorlinks  = true,
  citecolor   = blue,
  urlcolor    = red,
  linkcolor   = black
}
\newcommand{\hepth}[1]{{\tt
\href{http://www.arXiv.org/abs/hep-th/#1}{hep-th/#1}}}
\newcommand{\grqc}[1]{{\tt
\href{http://www.arXiv.org/abs/gr-qc/#1}{gr-qc/#1}}}

\newcommand{\arxiv}[1]{{\tt arXiv:\href{http://www.arXiv.org/abs/#1}{#1}}}
\newcommand{\doi}[1]{{\tt DOI:\href{http://dx.doi.org/#1}{#1}}}

\usepackage{tikz}\newcommand{\FPAUO}[2]{
\tikz[scale=.13,
         Uniovi/.style={color=green!51!blue, fill=green!51!blue}
 ] {
 \fill[Uniovi] (0,0) circle (10);
 \fill[white] (0,7) circle (1.5);
 \draw[Uniovi] (-2,7.5) rectangle (2,5.5);
 \fill[white] (-0.3,6.6) rectangle (0.3,0);   
 \fill[white] ( -0.9,6.2) rectangle (.9 ,5.6);
 \fill[white] (-1.4, 5.2) rectangle (1.4, 4.6);
 \fill[white] (0,0) ellipse (3.5 and 4);
 \fill[Uniovi] (-2.5,0.3) rectangle (2.5,-0.3);
 \fill[Uniovi] (-2,2.3) rectangle (2,1.7);
 \fill[Uniovi] (-2,-2.3) rectangle (2,-1.7);
 \fill[white] (-4.5,5.5) rectangle (-2.7,4.9);
 \fill[white] (-3.9,6.1) rectangle (-3.3,4.3);
 \fill[white] (4.5,5.5) rectangle (2.7,4.9);
 \fill[white] (3.9,6.1) rectangle (3.3,4.3);
 \foreach \x in { 0,..., 3 }
   \foreach \y in { 0,...,\x}
    {
     \fill[white] (-6-\x*0.7+\y*1.4,3.5-\x *1.97) -- (-5.6-\x*0.7+\y*1.4,2.4-\x *1.97) -- (-6.4-\x*0.7+\y*1.4,2.4-\x *1.97) -- cycle;
     \fill[white] (6-\x*0.7+\y*1.4,3.5-\x *1.97) -- (5.6-\x*0.7+\y*1.4,2.4-\x *1.97) -- (6.4-\x*0.7+\y*1.4,2.4-\x *1.97) -- cycle;
   };
 \draw (0,-6) node[
                               text centered, 
                               color=white, 
                               font={\fontsize{8}{4}\sffamily\selectfont}
                             ] {FPAUO-#1/#2};
}} 
\usepackage{pgfplots}    
\makeatletter
\@addtoreset{equation}{section}
\makeatother

\begin{document}

\vspace{-1cm}
\begin{flushright}
\small
\FPAUO{18}{01}\\
IFT-UAM/CSIC-17-107\\
IFUM-1057-FT\\
\texttt{arXiv:1803.01919 [hep-th]}\\
March 5\textsuperscript{th}, 2018\\
\normalsize
\end{flushright}

\vspace{0cm}

\begin{center}

{\Large {\bf {$\alpha'$-corrected black holes in String Theory}}}

\vspace{0.8cm}

\renewcommand{\thefootnote}{\alph{footnote}}
{\sl\large Pablo A.~Cano$^{1,2}$}${}^{,}$\footnote{E-mail: {\tt pablo.cano [at] uam.es}},
{\sl\large Patrick Meessen$^{3}$}${}^{,}$\footnote{E-mail: {\tt meessenpatrick [at] uniovi.es}},
{\sl\large Tom\'{a}s Ort\'{\i}n$^{1}$}${}^{,}$\footnote{E-mail: {\tt Tomas.Ortin [at] csic.es}}
{\sl\large and Pedro F.~Ram\'{\i}rez$^{4}$}${}^{,}$\footnote{E-mail: {\tt
    ramirez.pedro  [at] mi.infn.it}},

\setcounter{footnote}{0}
\renewcommand{\thefootnote}{\arabic{footnote}}

\vspace{0.5cm}

${}^{1}${\it Instituto de F\'{\i}sica Te\'orica UAM/CSIC\\
C/ Nicol\'as Cabrera, 13--15,  C.U.~Cantoblanco, E-28049 Madrid, Spain}\\ 
\vspace{0.2cm}

${}^{2}${\it Perimeter Institute for Theoretical Physics\\
  Waterloo, ON N2L 2Y5, Canada}\\
  \vspace{0.2cm}

${}^{3}${\it HEP Theory Group, Departamento de F\'{\i}sica, Universidad de Oviedo\\
  Avda.~Calvo Sotelo s/n, E-33007 Oviedo, Spain}\\
    \vspace{0.2cm}

${}^{4}${\it INFN, Sezione di Milano,\\ Via Celoria 16, 20133 Milano, Italy.
}\\

\vspace{.5cm}


{\bf Abstract}

\end{center}

\begin{quotation}
  {\small We consider the well-known solution of the Heterotic Superstring
    effective action to zeroth order in $\alpha'$ that describes the
    intersection of a fundamental string with momentum and a solitonic 5-brane
    and which gives a 3-charge, static, extremal, supersymmetric black hole in
    5 dimensions upon dimensional reduction on $\mathrm{T}^{5}$. We compute
    explicitly the first-order in $\alpha'$ corrections to this solution,
    including $\mathrm{SU}(2)$ Yang-Mills fields which can be used to cancel
    some of these corrections and we study the main properties of this
    $\alpha'$-corrected solution: supersymmetry, values of the near-horizon
    and asymptotic charges, behavior under $\alpha'$-corrected T-duality,
    value of the entropy (using Wald formula directly in 10 dimensions),
    existence of small black holes etc. The value obtained for the entropy
    agrees, within the limits of approximation, with that obtained by
    microscopic methods. 
    The $\alpha'$ corrections coming from Wald's formula prove
    crucial for this result.
}
\end{quotation}

\newpage
\pagestyle{plain}

\tableofcontents


\section*{Introduction}

Ever since Strominger and Vafa's computation of the microscopic entropy of an
extremal, static, 3-charge black hole in 5 dimensions
\cite{Strominger:1996sh}, showing perfect agreement at first order with the
macroscopic (Bekenstein-Hawking) entropy, there has been a keen interest in
going beyond this approximation both at the microscopic and macroscopic
levels.

Going beyond the first approximation at the macroscopic level involves
considering corrections to the superstring field theory effective action and
finding solutions of the corresponding equations of motion valid to the
required approximation level that describe black holes. Then one needs to use
an entropy formula such as Wald's \cite{Wald:1993nt,Iyer:1994ys} to take into
account the corrections to the action and not just the corrected geometry of
the solution.

Independently of their origin (string or worldsheet loops) the corrections to
the superstring effective action are terms of higher order in the curvatures
and take a very complicated form, specially after compactification. Thus, no
successful attempts to solving the corrected equations of motion for black
holes have been made so far and researchers in this field have adopted
different strategies to simplify the problem: either considering only a number
of tractable corrections (the Gauss-Bonnet term is one of them) which may appear
integrated in the structure (the prepotential) of an otherwise normal,
quadratic, $\mathcal{N}=2,d=4$ supergravity (see,
\textit{e.g.}~Ref.~\cite{Mohaupt:2000mj} and references therein) or by dealing
only with the near-horizon solution through different approaches (see
\textit{e.g.}~Ref.~\cite{Prester:2010cw} and references therein). 

In both cases it is argued that the most important corrections are being
captured, basically because the expected result is found, but a definite proof
is not available. Dealing with near-horizon geometries, for instance, leads to
the problem of finding the total, asymptotic charges of the black holes which
occur in the mass formula and some of the corrections to the entropy are
attributed to the difference between near-horizon and total charges which,
actually, are not known.  Furthermore, the calculation of the entropy is also
affected by the lack of knowledge of the complete action, even if the
near-horizon geometry is known (by hypothesis). 

The fact that, in general, the microscopic entropies are reproduced by these
methods can only be regarded as circumstantial evidence of their validity.
Only the explicit knowledge of the complete (near-horizon to infinity)
$\alpha'$-corrected black hole solutions and the subsequent calculation of the
entropy using the full action can clarify the situation.

In this paper we carry out this program for the same 3-charge 5-dimensional
black hole considered by Strominger and Vafa in the context of the Heterotic
Superstring effective action, to first order in $\alpha'$: we find the
explicit $\alpha'$ corrections to all the fields of the solution and then we
apply Wald's formula to the complete action obtaining an unambiguous answer
that reproduces the microscopic result found in Ref.~\cite{Castro:2008ys}.  As
we will show, this is possible because we carry out all the calculations
directly in 10 dimensions and, for these black-hole solutions all the
$\alpha'$ corrections are the Laplacian of a function which provides the
correction to the harmonic functions of the zeroth-order solution.

We have found it convenient to add a $\mathrm{SU}(2)$ instanton field to
Strominger and Vafa's solution because, as we will see, it can be used at
pleasure to make arbitrarily small or cancel identically many of the $\alpha'$
corrections. This cancellation takes place not just at the level of the field
strengths and curvatures, but also at the level of the Chern-Simons term via a
mechanism that we will explain in full detail in a coming publication
\cite{Chimento:2018kop}.

Since the corrections associated to the gauge fields have the same form as
those associated to the curvature of the torsionful spin connection, it also
helps us to better understand the latter and the nature of the so-called
\textit{symmetric 5-brane}, found in Ref.~\cite{Callan:1991dj} which is known
to be an exact solution of the Heterotic Superstring effective action to all
orders in $\alpha'$.

This paper is organized as follows: in Section~\ref{sec-heteroticalpha} we
review the Heterotic Superstring effective action, its fermionic supersymmetry
transformations and its equations of motion to $\mathcal{O}(\alpha')$. Since
most of this work will be carried out in 10-dimensional language, this section
sets the basis and the conventions for the rest of the paper. In
Section~\ref{sec-corrections} we propose a 10-dimensional ansatz for the
$\alpha'$-corrected solution that reduces to the Strominger and Vafa's
3-charge black hole when $\alpha'=0$, we show that it preserves 4 out of 16
supercherges (Section~\ref{sec-unbrokensusy}), plug it into the equations of
motion of the previous section, and solve for the undetermined functions.  In
Section~\ref{sec-chargecorrections} we start the study of the
$\alpha'$-corrected solution by computing the numbers of branes that source
the solution and trying to understand their relation with the total,
asymptotic charges of the fields. In order to gain a better understanding of
this point, in Section~\ref{sec-T-duality} we explore the behavior under
T-duality of this solution using the $\alpha'$-corrected Buscher T-duality
rules proposed in Ref.~\cite{Bergshoeff:1995cg}. Both the solution and the
T-duality rules pass the test.\footnote{In Ref.~\cite{Edelstein:2018ewc},
essentially the same $\alpha'$-corrected T-duality rules have been used to 
show the invariance of the temperature and entropy of the BTZ black hole 
in a simplified model.} In Section~\ref{sec-bhcorrections} we study the
$\alpha'$-corrected geometry of the 5-dimensional black hole that one obtains
by compactification of the solution on a $\mathrm{T}^{5}$. Finding the form of
all the 5-dimensional fields is very complicated (it requires performing the
compactification of the corrected action), except for the metric and the
dilaton, which are the 5-dimensional fields that interest us the most. This
allows us to find under which conditions there is a regular horizon and
compute the area of the horizon (the entropy of the zeroth-order solution) and
the mass of the solution. Then, in Section~\ref{sec-entropy} we compute the
corrections to the entropy using Wald's formula in 10-dimensional form. We
find two possible corrections to first-order in $\alpha'$, one of which
vanishes identically due to the very special properties of the 10-dimensional
near-horizon geometry \cite{Prester:2008iu}. The $\alpha'$-corrected entropy
reproduces the expected result once the difference in conventions have been
taken into account. In Section~\ref{sec-massless} we study the issue of the
existence of \textit{small black holes} with classical vanishing
area. Finally, in Section~\ref{sec-validity} we study the limits under which
the solution can be considered a good first-order in $\alpha'$ approximation
to an exact solution of the full Heterotic Superstring effective action.
Section~\ref{sec-discussion} contains our conclusions.

\section{The Heterotic Superstring effective action to
  $\mathcal{O}(\alpha')$}
\label{sec-heteroticalpha}

The Heterotic Superstring effective action to $\mathcal{O}(\alpha')$ can be
written in the string frame in the following concise form
\cite{Bergshoeff:1989de}:\footnote{We follow the conventions of
  Ref.~\cite{Ortin:2015hya} for the spin connection and curvature and for the
  gamma matrices. See also Ref.~\cite{Chemissany:2007he}.}

\begin{equation}
\label{heterotic}
{S}
=
\frac{g_{s}^{2}}{16\pi G_{N}^{(10)}}
\int d^{10}x\sqrt{|{g}|}\, 
e^{-2{\phi}}\, 
\left\{
{R} 
-4(\partial{\phi})^{2}
+\tfrac{1}{2\cdot 3!}{H}^{2}
-\tfrac{1}{2}T^{(0)}
\right\}\, .
\end{equation}

\noindent
Let us now review the definition of the different terms that appear in
it. First of all, $\phi$ is the dilaton field and the vacuum expected value of
$e^{\phi}$ is the Heterotic Superstring coupling constant $g_{s}$. The
10-dimensional Newton constant $G_{N}^{(10)}$ is given in terms the string
length $\ell_{s}$ (with $\alpha'=\ell_{s}^{2}$) and $g_{s}$
by

\begin{equation}
\label{eq:d10newtonconstant}
G_{N}^{(10)}=8\pi^{6}g_{s}^{2} \ell_{s}^{8}\, .
\end{equation}

\noindent
$R$ is the Ricci scalar of the string-frame metric $g_{\mu\nu}$. $T^{(0)}$ is
one of the three ``$T$-tensors'' associated to $\alpha'$ corrections and which
are defined as

\begin{equation}
\label{eq:Ttensors}
\begin{array}{rcl}
{T}^{(4)}
& \equiv &
6
\alpha'\left[
{F}^{A}\wedge{F}^{A}
+
{R}_{(-)}{}^{{a}}{}_{{b}}
\wedge
{R}_{(-)}{}^{{b}}{}_{{a}}
\right]\, ,
\\
& & \\ 
{T}^{(2)}{}_{{\mu}{\nu}}
& \equiv &
2\alpha'\left[
{F}^{A}{}_{{\mu}{\rho}}{F}^{A}{}_{{\nu}}{}^{{\rho}} 
+
{R}_{(-)\, {\mu}{\rho}}{}^{{a}}{}_{{b}}
{R}_{(-)\, {\nu}}{}^{{\rho}\,  {b}}{}_{{a}}
\right]\, ,
\\
& & \\    
{T}^{(0)}
& \equiv &
{T}^{(2)\,\mu}{}_{{\mu}}\, .
\\
\end{array}
\end{equation}

\noindent
In these definitions, ${R}_{(-)}{}^{{a}}{}_{{b}}$ is one of the two Lorenz
curvature 2-forms ${R}_{(\pm)}{}^{{a}}{}_{{b}}$ of the two \textit{torsionful
  spin connection 1-forms} ${\Omega}_{(\pm)}{}^{{a}}{}_{{b}}$ that can be
constructed by combining the Levi-Civita spin connection ${\omega}^{{a}{b}}$
1-form with a torsion piece proportional to the Kalb-Ramond field strength
$H$. ${F}^{A}$ is the $\mathrm{SU}(2)$ Yang-Mills field strength and ${H}$ is the
Kalb-Ramond field strength 3-form.  All these objects are defined by

\begin{eqnarray}
{\Omega}_{(\pm)}{}^{{a}}{}_{{b}} 
& = &
{\omega}^{{a}}{}_{{b}}
\pm
\tfrac{1}{2}{H}_{{\mu}}{}^{{a}}{}_{{b}}dx^{{\mu}}\, ,
\\
& & \nonumber \\
{R}_{(\pm)}{}^{{a}}{}_{{b}}
& = & 
d {\Omega}_{(\pm)}{}^{{a}}{}_{{b}}
- {\Omega}_{(\pm)}{}^{{a}}{}_{{c}}\wedge  
{\Omega}_{(\pm)}{}^{{c}}{}_{{b}}\, ,
\\
& & \nonumber \\
{F}^{A}
& =&
d{A}^{A}+\tfrac{1}{2}\epsilon^{ABC}{A}^{B}\wedge{A}^{C}\, , 
\\
& & \nonumber \\
\label{Hdef}
{H}
& = & 
d{B}
+2\alpha'\left({\omega}^{\rm YM}
+{\omega}^{{\rm L}}_{(-)}\right)\, .
\end{eqnarray}

\noindent
In the definition of $H$, ${\omega}^{\rm YM}$ and ${\omega}^{{\rm L}}_{(-)}$
are, respectively, the Yang-Mills and Lorentz Chern-Simons terms

\begin{eqnarray}
{\omega}^{\rm YM}
& = & 
dA^{A}\wedge {A}^{A}
+\tfrac{1}{3}\epsilon^{ABC}{A}^{A}\wedge{A}^{B}\wedge{A}^{C}\, ,
\\
& & \nonumber \\
{\omega}^{{\rm L}}_{(\pm)}
& = & 
d{\Omega}_{(\pm)}{}^{{a}}{}_{{b}} \wedge 
{\Omega}_{(\pm)}{}^{{b}}{}_{{a}} 
-\tfrac{2}{3}
{\Omega}_{(\pm)}{}^{{a}}{}_{{b}} \wedge 
{\Omega}_{(\pm)}{}^{{b}}{}_{{c}} \wedge
{\Omega}_{(\pm)}{}^{{c}}{}_{{a}}\, .
\end{eqnarray}

\noindent
Then, the Bianchi identity of $H$ is 

\begin{equation}
\label{eq:BianchiH}
d{H}  
-
\tfrac{1}{3}T^{(4)}
=
0\, .
\end{equation}

The above action contains an infinite number of implicit $\alpha'$ corrections
which arise due to the recursive way in which ${H}$ is defined: $H$ depends on
the Lorentz Chern-Simons form of ${\omega}^{L}_{(-)}$, which depends on
${\Omega}_{(-)}$, which, in its turn, is defined in terms of ${H}$. At the
order at which we are working, it is enough to keep in the definitions of
${\Omega}_{(\pm)}$ only the terms of zeroth order in $\alpha'$, that is

\begin{equation}
{\Omega}_{(\pm)}{}^{{a}}{}_{{b}} 
=
{\omega}^{{a}}{}_{{b}}
\pm
\tfrac{1}{2}{H}^{(0)}_{{\mu}}{}^{{a}}{}_{{b}}dx^{{\mu}}\, ,
\,\,\,\,\,
\mbox{where}
\,\,\,\,\,
{H}^{(0)} \equiv d{B}\, .
\end{equation}

\noindent
Furthermore we will ignore all the $\alpha'^2$ terms in the action.

The equations of motion that follow from this action are very complicated and,
in order to deal with them, we proceed as in Section~3 of
Ref.~\cite{Bergshoeff:1992cw}: we separate the variations with respect to each
field ($g_{\mu\nu}, B_{\mu\nu},\phi,A^{A}_{\mu}$) into those corresponding to
the \textit{explicit} occurrences of the fields in the action
(\textit{i.e.}~when they do not appear in ${\Omega}_{(-)}{}^{{a}}{}_{{b}}$)
and those corresponding to \textit{implicit} occurrences via
${\Omega}_{(-)}{}^{{a}}{}_{{b}}$:

\begin{eqnarray}
\delta S 
& = &  
\frac{\delta S}{\delta g_{\mu\nu}}\delta g_{\mu\nu}
+\frac{\delta S}{\delta B_{\mu\nu}}\delta B_{\mu\nu}
+\frac{\delta S}{\delta A^{A}{}_{\mu}}\delta A^{A}{}_{\mu}
+\frac{\delta S}{\delta \phi} \delta \phi
\nonumber \\
& & \nonumber \\
& = & 
\left.\frac{\delta S}{\delta g_{\mu\nu}}\right|_{\rm exp.}\delta g_{\mu\nu}
+\left.\frac{\delta S}{\delta B_{\mu\nu}}\right|_{\rm exp.}\delta B_{\mu\nu}
+\left.\frac{\delta S}{\delta A^{A}{}_{\mu}}\right|_{\rm exp.}\delta A^{A}{}_{\mu}
+\frac{\delta S}{\delta \phi} \delta \phi
\nonumber \\
& & \nonumber \\
& &
+\frac{\delta S}{ \delta {\Omega}_{(-)}{}^{{a}}{}_{{b}}}
\left(
\frac{\delta {\Omega}_{(-)}{}^{{a}}{}_{{b}}}{\delta g_{\mu\nu}}\delta g_{\mu\nu}
+\frac{\delta {\Omega}_{(-)}{}^{{a}}{}_{{b}}}{\delta B_{\mu\nu}} \delta B_{\mu\nu}
+\frac{\delta {\Omega}_{(-)}{}^{{a}}{}_{{b}}}{\delta A^{A}{}_{\mu}}\delta
A^{A}{}_{\mu}
\right)\, .
\end{eqnarray}

\noindent
Written in this way, we can then make use of the lemma proven in Section~3 of
Ref.~\cite{Bergshoeff:1989de}: $\delta S/\delta
{\Omega}_{(-)}{}^{{a}}{}_{{b}}$ is proportional to $\alpha'$ and to the
zeroth-order equations of motion of $g_{\mu\nu},B_{\mu\nu}$ and $\phi$ plus
terms of higher order in $\alpha'$. Thus, for any solution of the zeroth-order
equations which is exact or up to terms of order $\alpha'$, these terms are,
at least, of order $\alpha'^{2}$ and can be safely ignored for our purposes.

The variations with respect to the explicit occurrences of the fields are,
after some manipulations

\begin{eqnarray}
\label{eq:eq1}
R_{\mu\nu} -2\nabla_{\mu}\partial_{\nu}\phi
+\tfrac{1}{4}{H}_{\mu\rho\sigma}{H}_{\nu}{}^{\rho\sigma}
-T^{(2)}{}_{\mu\nu}
& = & 
0\, ,
\\
& & \nonumber \\
\label{eq:eq2}
(\partial \phi)^{2} -\tfrac{1}{2}\nabla^{2}\phi
-\tfrac{1}{4\cdot 3!}{H}^{2}
+\tfrac{1}{8}T^{(0)}
& = &
0\, ,
\\
& & \nonumber \\
\label{eq:eq3}
d\left(e^{-2\phi}\star {H}\right)
& = &
0\, ,
\\
& & \nonumber \\
\label{eq:eq4}
\alpha' e^{2\phi}\mathfrak{D}_{(+)}\left(e^{-2\phi}\star {F}^{A}\right)
& = & 
0\, ,
\end{eqnarray}

\noindent
where $\mathfrak{D}_{(+)}$ is the exterior derivative covariant with respect
to the $\mathrm{SU}(2)$ group and with respect to the torsionful connection
$\Omega_{(+)}$, that is

\begin{equation}
e^{2\phi}d\left(e^{-2\phi}\star {F}^{A}\right)
+\epsilon^{ABC}{A}^{B}\wedge \star F^{C}
+\star {H}\wedge{F}^{A}
= 
0\, .   
\end{equation}

\noindent
The three non-trivial zeroth-order equations can be obtained from these by
setting $\alpha'=0$. This eliminates the Yang-Mills fields, the $T$-tensors
and the Chern-Simons terms in $H$.

We are also going to need the supersymmetry transformation laws of the
gravitino $\psi_{\mu}$, dilatino $\lambda$ and gaugini $\chi^{A}$ for
vanishing fermions, to find the unbroken supersymmetries of the field
configurations under study. These are given by

\begin{eqnarray}
\label{eq:deltapsi}
\delta_{\epsilon}\psi_{\mu}
& = &
\nabla_{\mu}^{(+)}\epsilon 
\equiv 
\left(
\partial_{\mu} 
-\tfrac{1}{4}\!\!\not\!\!\Omega_{(+)\, \mu}
\right)\epsilon\, ,
\\
& & \nonumber \\
\label{eq:deltalambda}
\delta_{\epsilon}\lambda
& = & 
\left(
\not\!\partial\phi -\tfrac{1}{12}\!\!\not\!\!H
\right)\epsilon\, ,
\\
& & \nonumber \\
\label{eq:deltachi}
\alpha' \delta_{\epsilon}\chi^{A}
& = & 
-\tfrac{1}{4}\alpha'\!\!\not\! F^{A} \epsilon\, .
\end{eqnarray}

\noindent
In these expressions $H$ includes the Chern-Simons terms, which provide the
first $\alpha'$ corrections.

\section{$\alpha'$ corrections to the $d=10$ Heterotic Superstring background}
\label{sec-corrections}

We are interested in the following 10-dimensional field configuration

\begin{equation}
\label{10dmetric}
\begin{array}{rcl}
d{s}^{2}
& = &
{\displaystyle
\frac{2}{\mathcal{Z}_{-}}du\left(dv-\tfrac{1}{2}\mathcal{Z}_{+}du\right)
-\mathcal{Z}_{0}(d\rho^{2}+\rho^{2}d\Omega_{(3)}^{2})-dy^{i}dy^{i}\, ,
\hspace{.5cm}
i=1,\ldots,4\, ,
}
\\
& & \\
H
& = & 
{\displaystyle
d\mathcal{Z}_{-}^{-1}\wedge du\wedge dv
-\frac{\rho^{3}{\mathcal{Z}_{0}'}}{8} \sin{\theta}
d\theta\wedge d\psi\wedge d\phi\, ,
}
\\
& & \\
{A}^{A}
& = & 
{\displaystyle
-\frac{\rho^{2}}{(\kappa^{2}+\rho^{2})}v^{A}_{L}\, ,
}
\\
& & \\
e^{-2{\phi}}
& = &
{\displaystyle
e^{-2{\phi}_{\infty}}\frac{\mathcal{Z}_{-}}{\mathcal{Z}_{0}}\, ,
}
\end{array}
\end{equation}

\noindent
where the functions $\mathcal{Z}_{+,-,0}$ are assumed to be of the form

\begin{equation}
\mathcal{Z}_{0}
=
1+\frac{\mathcal{Q}_{0}}{\rho^{2}}+\alpha' f_{0}(\rho)\, ,
\hspace{1.5cm}
\mathcal{Z}_{\pm}
=
1+\frac{Q_{\pm}}{\rho^{2}}+\alpha' f_{\pm}(\rho)\, ,
\end{equation}

\noindent
where, in their turn, $f_{\pm,0}$ are functions of $\rho$ to be
determined. Observe that all the functions in this ansatz depend on the radial
coordinate $\rho$ of a $\mathbb{R}^{4}$ space, which is adequate for single,
static, branes and black holes. The connections and curvatures for this ansatz
are computed in Appendix~\ref{sec-connection} in a slightly more general form,
using Cartesian coordinates $x^{m}$ with $x^{m}x^{m}=\rho^{2}$. 

When the undetermined functions $f_{+,-,0}$ and the $\mathrm{SU}(2)$ gauge
field are set to zero, this field configuration is a well known $1/4$
supersymmetric solution of the zeroth-order equations of the Heterotic
Superstring effective action \cite{Tseytlin:1996as,Cvetic:1995bj} describing
an intersection or superposition of

\begin{enumerate}

\item \textit{Solitonic (S)} or \textit{Neveu-Schwarz (NS)} 5-branes
  \cite{Duff:1990wv,Callan:1991dj},\footnote{We feel more inclined to use the
    name S5-branes.} lying in the directions $u,v,y^{1},\cdots,y^{4}$. The
  $\mathbb{R}^{4}$ space parametrized by the coordinates $x^{m}$,
  $m=1,\cdots,4$ is their transverse space and it is the common transverse
  space of the whole solution. They are described by the function
  $\mathcal{Z}_{0}$ and their charge is represented by $\mathcal{Q}_{0}$ at
  this order.

\item A \textit{fundamental string (F1)} lying in the directions $u,v$ and
  smeared over the rest of the S5-branes' worldvolume directions $y^{i}$,
  $i=1,\cdots,4$. It is described by the function $\mathcal{Z}_{-}$ and its
  charge (winding number) is represented by $\mathcal{Q}_{-}$ at this order.

\item A \textit{gravitational pp-wave (W)} carrying momentum along the $v$
  direction (\textit{i.e.}~along the F1). It is described by the function
  $\mathcal{Z}_{-}$ and its charge (momentum) is represented by
  $\mathcal{Q}_{+}$ at this order. The interchange between $\mathcal{Q}_{+}$
  and $\mathcal{Q}_{-}$ under T-duality at zeroth order in $\alpha'$
  corresponds to the interchange between winding and momentum of the F1.

\end{enumerate}

Upon dimensional reduction over a $\mathrm{T}^{5}$, this solution gives a
static,extremal, 3-charge, $1/2$ supersymmetric black hole in
$\mathcal{N}=1,d=5$ supergravity, which is dual to the one studied by
Strominger and Vafa in Ref.~\cite{Strominger:1996sh}.\footnote{See also
  Refs.~\cite{Maldacena:1996ky,Peet:2000hn,David:2002wn}.}

In Ref.~\cite{Cano:2017qrq} we considered the addition of the above
$\mathrm{SU}(2)$ gauge field, which is nothing but a BPST instanton, in the
context of Heterotic Supergravity. Heterotic Supergravity is just
$\mathcal{N}=1,d=10$ supergravity coupled to vector supermultiplets and can be
obtained from the Heterotic Superstring effective action in
Eq.~(\ref{heterotic}) by eliminating all the terms containing the torsionful
spin connection $\Omega_{(-)}$.  Thus, it only contains part of the $\alpha'$
terms of the Heterotic Superstring effective action. However, it is exactly
invariant under supersymmetry \cite{Bergshoeff:1989de}, which makes it easier
to use supersymmetric solution-generating techniques and, indeed, using these
techniques in $\mathcal{N}=1,d=5$ gauged supergravity it was shown that 
with $f_{0}$ given by

\begin{equation}
\label{eq:f0exactsolution}
f_{0}(\rho)
=
8\frac{\rho^{2} +2\kappa^{2}}{(\rho^{2} +\kappa^{2})^{2}}\, ,  
\end{equation}

\noindent
and $f_{+}=f_{-}=0$ the above field configuration is an exact supersymmetric
solution which, upon dimensional reduction over a $\mathrm{T}^{5}$ gives a
static, extremal, 3-charge, $1/2$ supersymmetric black hole in
$\mathcal{N}=1,d=5$ supergravity with non-Abelian hair
\cite{Bellorin:2007yp,Bueno:2015wva,Meessen:2015enl,Cano:2017qrq}.

To be more precise, $f_{0}(\rho)$ is defined up to an arbitrary harmonic
function. In Eq.~(\ref{eq:f0exactsolution}) the harmonic function has been
chosen so as to make $f_{0}(\rho)$ regular at $\rho=0$ while keeping the
normalization of $\mathcal{Z}_{0}$ at infinity. We will always use the same
convention to choose the arbitrary harmonic functions that can be added to
$f_{0,+,-}(\rho)$.  With this convention, the only $1/\rho^{2}$ pole in
$\rho\to 0$ limit of the $\alpha'$-corrected $\mathcal{Z}_{0}$ is the original
term $\mathcal{Q}_{0}/\rho^{2}$ where $\mathcal{Q}_{0}$ is proportional to the
number of S5-branes \cite{Duff:1990wv,Callan:1991dj}.

Further shifts by harmonic functions can always be absorbed into a
redefinition of $\mathcal{Q}_{0}$ Observe that, in the $\rho\to \infty$ limit,
the coefficient of the $1/\rho^{2}$ term is not $\mathcal{Q}_{0}$ but
$\mathcal{Q}_{0}+8\alpha'$. The difference is due to the contribution of the
BPST instanton which sources a ``gauge 5-brane''
\cite{Strominger:1990et,Cano:2017sqy} which in its turn increases to the total
charge of the NS 6-form $\tilde{B}$ dual to the Kalb-Ramond 2-form $B$
measured at infinity. In this case, the difference between these two
quantities, number of S5-branes and total 5-brane charge at infinity, has a
simple explanation in terms of a delocalized gauge 5-brane but, as we are
going to see, other $\alpha'$ corrections lead to very similar differences
between ``near-horizon'' and ``asymptotic'' (total) charges which do not have
a (known) similar, simple, interpretation.

The fact that the $\alpha'$ corrections associated to the torsionful spin
connection $\Omega_{(-)}$ have the same structure as those associated to the
gauge fields should not come as a surprise: the theory treats the Yang-Mills
and the torsionful spin connection on exactly the same footing
\cite{Bergshoeff:1988nn} and the curvature of the latter occurs as that of
another non-Abelian gauge field sourcing the Einstein equations. The main
difference is that the torsionful spin connection is not an independent field
and, furthermore, its ``kinetic term'' occurs in the action with the wrong
sign.

Thus, on general grounds, one expects additional $\alpha'$ corrections in
$f_{+,-,0}(\rho)$ similar to Eq.~(\ref{eq:f0exactsolution}), with opposite
sign and depending on $\mathcal{Q}_{+},\mathcal{Q}_{-},\mathcal{Q}_{0}$
instead of $\kappa$. These corrections cannot be assigned to something like a
``gravitational 5-brane'', as far as we know, but they are similarly
delocalized and they will generically contribute to the total charges at
infinity. This may give rise to the problem of how to count the number of
branes through the computation of the charge.

Remarkably enough, when the instanton field is included together with the rest
of first-order $\alpha'$ corrections, some of of these contributions to the
total charge disappear completely and the total charge at infinity has the
same value as the ``near-horizon'' ($\rho\to 0$) charge, as it happens at
zeroth order in $\alpha'$. Actually, since, according to the previous
discussion, the structure of those corrections is the same as that of those
associated to the Yang-Mills fields we can cancel them against each other,
eliminate completely the first-order $\alpha'$ corrections and (probably, we
conjecture) all the higher order corrections.  It is likely that the addition
of more general gauge fields can be used to solve this problem for all charges
and also to, eventually, cancel all the $\alpha'$ corrections
\cite{kn:CChMORR}.

We will discuss this issue at length in Section~\ref{sec-chargecorrections}.

Right now our goal is to determine the functions $f_{+,-,0}(\rho)$ so that the
above field configuration is a solution of the Heterotic Superstring effective
action to first order in $\alpha'$ (\textit{i.e.}~up to terms of
$\mathcal{O}(\alpha'^{2})$). However, before doing it, we are going to show
that these field configurations preserve $1/4$ of the supersymmetries for any
value of the functions $\mathcal{Z}_{+},\mathcal{Z}_{-},\mathcal{Z}_{0}$ and
for any Yang-Mills field strength which is self-dual in the 4-dimensional
space $\mathbb{R}^{4}$ transverse to the S5-branes to first order in
$\alpha'$.

\subsection{Unbroken supersymmetries of the ansatz}
\label{sec-unbrokensusy}

Using the Zehnbein basis and results in Appendix~\ref{sec-connection} for the
torsionful spin connection $\Omega_{(+)}$, the different components of the
supersymmetry transformation rules
Eqs.~(\ref{eq:deltapsi})-(\ref{eq:deltachi}) take the following form for our
ansatz:

\begin{eqnarray}
\delta_{\epsilon}\psi_{+}
& = &
\left[
\partial_{+} 
+\tfrac{1}{4}
\frac{\mathcal{Z}_{-}\partial_{m}\mathcal{Z}_{+}}{\mathcal{Z}_{0}^{1/2}}
\Gamma^{m}\Gamma^{+}
\right]\epsilon\, ,
\\
& & \nonumber \\
\delta_{\epsilon}\psi_{-}
& = &
\left[
\partial_{-} 
+\tfrac{1}{4}\frac{\partial_{m}\log{\mathcal{Z}_{-}}}{\mathcal{Z}_{0}^{1/2}}
\Gamma^{m}\Gamma^{+}
\right]\epsilon\, ,
\\
& & \nonumber \\
\delta_{\epsilon}\psi_{m}
& = &
\left[
\partial_{m} 
+\tfrac{1}{8}\frac{\partial_{q}\log{\mathcal{Z}_{0}}}{\mathcal{Z}_{0}^{1/2}} 
(\mathbb{M}^{+}_{qm})_{np}\Gamma^{np}(1-\tilde{\Gamma})
\right]\epsilon\, ,
\\
& & \nonumber \\
\delta_{\epsilon}\psi_{i}
& = &
\partial_{i} 
\epsilon\, ,
\\
& & \nonumber \\
\delta_{\epsilon}\lambda
& = & 
-\frac{1}{2\mathcal{Z}_{0}^{1/2}}\Gamma^{m}
\left[
\partial_{m}\log{\mathcal{Z}_{-}}\Gamma^{-}\Gamma^{+}
-\partial_{m}\log{\mathcal{Z}_{0}} (1-\tilde{\Gamma})
\right]\epsilon\, ,
\\
& & \nonumber \\
\alpha' \delta_{\epsilon}\chi^{A}
& = & 
-\frac{1}{8\mathcal{Z}_{0}^{1/2}}\alpha'\!\!\not\! F^{A} 
(1-\tilde{\Gamma})\epsilon\, ,
\end{eqnarray}

\noindent
where $\tilde{\Gamma}\equiv\Gamma^{2345}$ is the chirality matrix in the
$\mathbb{R}^{4}$ space transverse to the S5-branes. All these transformations
vanish identically for constant spinors satisfying the constraints

\begin{equation}
\tilde{\Gamma}\epsilon=+\epsilon\, ,
\hspace{1cm}
\Gamma^{+}\epsilon=0\, ,
\end{equation}

\noindent
which reduce the number of independent components to $1/4$ of the 16.

\subsection{Explicit computation of the $\alpha'$ corrections}

We just have to plug the supersymmetric configuration Eq.~(\ref{10dmetric}) in
the equations of motion (\ref{eq:eq1})-(\ref{eq:eq4}) as well as in the
Bianchi identity Eq.~(\ref{eq:BianchiH}) and try to solve them for
$f_{+,-,0}(\rho)$.  Our ansatz assumes implicitly that no more components of
the metric are necessary to this order and that its structure and symmetries
will remain intact. Only the functions associated to the different branes can
receive corrections. These assumptions based in our experience with the
non-Abelian black hole of Ref.~\cite{Cano:2017qrq} will prove correct, as we
are going to see.

The terms of order $\alpha'$ in Eqs.~(\ref{eq:eq1})-(\ref{eq:eq4}) are
proportional to the $T$-tensors defined in Eq.~(\ref{eq:Ttensors}), which were
computed for this ansatz in Ref.~\cite{Cano:2017qrq}. They are explicitly
given by\footnote{$\hat{T}^{(4)}$ is computed explicitly in
  Appendix~\ref{sec-connection}.}

\begin{eqnarray}
\label{eq:T4}
\hat{T}^{(4)}
& \sim & 
\alpha'\left[\frac{\kappa^{4}}{(\kappa^{2}+\rho^{2})^{4}}
-\frac{\mathcal{Q}_{0}^{2}}{(\mathcal{Q}_{0}+\rho^{2})^{4}}\right]
d\rho \rho^{3}\wedge \sin{\theta} d\theta\wedge d\Psi\wedge d\phi
\, ,
\\
& & \nonumber \\
\label{eq:T2uu}
\hat{T}^{(2)}{}_{uu}
& = &
- \alpha'\frac{32 \mathcal{Q}_{-} \mathcal{Q}_{+} \rho^{4} \left[\mathcal{Q}_{0}^{2}
+\mathcal{Q}_{0} \left(\mathcal{Q}_{-}+3 \rho^{2}\right)
+\mathcal{Q}_{-}^{2}+3 \mathcal{Q}_{-} \rho^{2}
+3\rho^{4}\right]}{\left(\mathcal{Q}_{0}+\rho^{2}\right)^{4} 
\left(\mathcal{Q}_{-}+\rho^{2}\right)^{4}}\, ,
\\
& & \nonumber \\
\label{eq:T2ij}
\hat{T}^{(2)}{}_{ij}
& = &
 \alpha'\delta_{ij}\frac{48 \rho^{2} }{\left(\mathcal{Q}_{0}+\rho^{2}\right)^5} 
\left[\mathcal{Q}_{0}^{2}-\frac{\kappa^{4} \left(\tilde
Q_{0}+\rho^{2}\right)^{4}}{\left(\kappa^{2}+\rho^{2}\right)^{4}}\right]\, ,
\\
& & \nonumber \\
\label{eq:T}
\hat{T}
& = &
- \alpha' \frac{192 \rho^{4} 
}{\left(\kappa^{2}+\rho^{2}\right)^{4}
\left(\mathcal{Q}_{0}+\rho^{2}\right)^{6}}
\left[\kappa^{8} \mathcal{Q}_{0}^{2}
+4 \kappa^{6} \mathcal{Q}_{0}^{2} \rho^{2}
\right.
\nonumber \\
& & \nonumber \\
& & 
\left.
-\kappa^{4} \left(\mathcal{Q}_{0}^{4}+4 \mathcal{Q}_{0}^{3}
\rho^{2}+4 \mathcal{Q}_{0} \rho^{6}+\rho^{8}\right)
+4 \kappa^{2} \mathcal{Q}_{0}^{2}
\rho^{6} +\mathcal{Q}_{0}^{2} \rho^8\right]\, .
\end{eqnarray}

Observe that, while all the scalar invariants that one can construct with
these $T$-tensors, and which occur in the action, depend on the parameters
$\mathcal{Q}_{0}$ and $\kappa^{2}$ only, the component $\hat{T}^{(2)}{}_{uu}$,
which occurs in the equations of motion, depends on
$\mathcal{Q}_{+},\mathcal{Q}_{-}$ and $\mathcal{Q}_{0}$ but not on
$\kappa^{2}$. $\hat{T}^{(2)}{}_{uu}$ vanishes identically at $\rho=0$, where
we expect the horizon to be, and it also vanishes asymptotically at $\rho\to
\infty$, but it is relevant at finite values of $\rho$. Thus, arguments solely
based on the behavior of the scalar invariants as functions of
$\mathcal{Q}_{0}$ and $\kappa^{2}$ miss completely this
correction. Furthermore, this correction disappears if one considers
near-horizon geometries only. 

Let us consider, first, the Yang-Mills fields. It can be seen that, given the
structure of the fields in our ansatz, independently of the actual values of
the $\mathcal{Z}$-functions, the $\alpha'$-corrected Yang-Mills equation
Eq.~(\ref{eq:eq4}) is satisfied automatically provided that ${F}^{A}$ is
self-dual in the 4-dimensional Euclidean space transverse to the S5-branes,
that is, $\star_{(4)}{F}^{A}=+{F}^{A}$, which is a property of our ansatz.

Next, we consider the Bianchi identity Eq.~(\ref{eq:BianchiH}) for the 3-form
$H$ in Eq.~(\ref{10dmetric}). Substituting the ansatz the identity takes the
form\footnote{We do not simplify the common factors in the left- and
  right-hand sides.} 

\begin{eqnarray}
-\frac{\alpha'}{8\rho^{3}}\frac{d}{d\rho}
\left(\rho^{3} \frac{df_{0}}{d\rho}\right)
d\rho \rho^{3}\wedge \sin{\theta} d\theta\wedge d\Psi\wedge d\phi
& = & 
\nonumber \\
& &  
\label{eq:equationforf0}\\
& & 
\hspace{-7.5cm}
24\alpha'
\left[
\frac{\kappa^{4}}{(\kappa^{2}+\rho^{2})^{4}}
-
\frac{\mathcal{Q}_{0}^{2}}{(\mathcal{Q}_{0}+\rho^{2})^{4}}
\right]
d\rho \rho^{3}\wedge \sin{\theta} d\theta\wedge d\Psi\wedge d\phi+\mathcal{O}(\alpha'^{2})\, .
\nonumber 
\end{eqnarray}

\noindent
This leads to the following equation for $f_{0}$

\begin{equation}
\frac{d}{d\rho}\left(\rho^{3} \frac{df_{0}}{d\rho}\right)
=
-192\rho^{3}\left[\frac{\kappa^{4}}{(\kappa^{2}+\rho^{2})^{4}}
-\frac{\mathcal{Q}_{0}^{2}}{(\mathcal{Q}_{0}+\rho^{2})^{4}}\right]\, ,
\end{equation}

\noindent
that can be integrated immediately, giving\footnote{This result is obtained in
  Appendix~\ref{sec-bianchiH} in a more transparent way. The integrability of
  this equation is due to a set of very interesting properties of this class
  of ansatzs that will be explored in more generality in Ref.~\cite{Chimento:2018kop}.}

\begin{equation}
\label{eq:f0solution}
f_{0}
=
8\left[
\frac{\rho^{2}+2\kappa^{2}}{(\rho^{2}+\kappa^{2})^{2}}
-
\frac{\rho^{2}+2\mathcal{Q}_{0}}{(\rho^{2}+\mathcal{Q}_{0})^{2}}\right]
+\frac{c_{0}}{\rho^{2}} +d_{0}\, .
\end{equation}

\noindent
Here $c_{0}$ and $d_{0}$ are integration constants corresponding to the
arbitrary shift by a harmonic function of $\mathcal{Z}_{0}$ discussed at the
beginning of this section. There we also convened to choose $c_{0}$ so that
$f_{0}$ has no $1/\rho^{2}$ poles in the $\rho\to 0$ limit and $d_{0}$ so that
$f_{0}$ vanishes asymptotically to preserve the asymptotic normalization of
the full metric. This has already been done in the expression above, which is
finite in the $\rho\to 0$ limit and vanishes in the $\rho\to\infty$
limit. Therefore,  $c_{0}=d_{0}=0$

Observe that, if $\mathcal{Q}_{0}=0$, the second term in
Eq.~(\ref{eq:f0solution}) should not be there at all. However, the above
expression gives a spurious $-1/\rho^{2}$ pole when $\mathcal{Q}_{0}=0$. Thus,
we will have to treat the cases $\mathcal{Q}_{0}=0$ and $\mathcal{Q}_{0}\neq
0$ independently. The same is also true for the $\kappa=0$ case, since in this
limit the instanton just gives an Abelian-like contribution that can be
interpreted as 8 S5-branes.\footnote{The Yang-Mills field becomes pure gauge
  in this limit except at $\rho=0$.}  Then we may simply reabsorb these 8
additional S5-branes into $\mathcal{Q}_{0}$.

The first term in $f_{0}$ is just the one in Eq.~(\ref{eq:f0exactsolution})
and is associated to the $F^{A}\wedge F^{A}$ term. The second is associated to
the ${R}_{(-)}{}^{{a}}{}_{{b}} \wedge {R}_{(-)}{}^{{b}}{}_{{a}} $ term and has
exactly the same structure because, as we said, $\Omega_{(-)}$ behaves exactly
as another gauge field. The presence of two terms with the same structure but
opposite signs ensures that the coefficient of the $1/\rho^{2}$ pole in the
$\rho\to 0$ limit is the same as the coefficient of the $1/\rho^{2}$ term in
the $\rho\to \infty$ limit: the contribution of the gauge 5-brane to the
charge of $\tilde{B}$ is cancelled by another contribution which cannot be
assigned to any known brane. Typically, the latter is the only $\alpha'$
correction considered in the literature in the context of black holes, where
uysually non-Abelian fields are not introduced.

The presence of two corrections with the same functional form but opposite
signs not only suppresses the difference between ``near-horizon'' and
asymptotic, total charge of $\tilde{B}$: setting $\kappa^{2}=\mathcal{Q}_{0}$
the whole first-order $\alpha'$ correction vanishes identically. With this
identification between the instanton size parameter and the S5-brane charge,
the component of the full solution described by $\mathcal{Z}_{0}$ is nothing
but the so-called \textit{symmetric 5-brane}, found in
Ref.~\cite{Callan:1991dj}, which is known to be an exact solution of the
Heterotic Superstring effective action to all orders in $\alpha'$.  Finding
the symmetric 5-brane solution in this form sheds new light on its origin and
meaning. Of course, the complete solution has additional fields which give
rise to some $\alpha'$ corrections of their own even if
$\kappa^{2}=\mathcal{Q}_{0}$, via $\hat{T}^{(2)}{}_{uu}$.

Finally, notice that if $\mathcal{Q}_{0}>>\kappa^{2}$ the second term is
irrelevant compared to the first one, except in the asymptotic limit
$\rho\rightarrow\infty$, where both are comparable. In
Ref.~\cite{Cano:2017qrq} this fact was used to argue that the solution of
Heterotic Supergravity that includes the instanton suffered only small
$\alpha'$ corrections to first order. In Section~\ref{sec-validity} we will
study the issue of $\alpha'$ and other corrections from a more general point
of view.

Next, let us consider the equation of motion of $B$, Eq.~(\ref{eq:eq3}). It
yields the following equation for $f_{-}$

\begin{equation}
\label{eq:equationforf-}
\frac{d}{d\rho}\left(\rho^{3} \frac{df_{-}}{d\rho}\right)=0\, ,
\end{equation}

\noindent
which means that $f_{-}(\rho)$ is just a harmonic function, which we absorb
into a redefinition of $\mathcal{Q}_{-}$ according to our general
prescription.  Therefore $\mathcal{Z}_{-}$ does not receive any first-order
$\alpha'$ corrections.

Now we can turn our attention to the Einstein equations Eq.~(\ref{eq:eq1}). We
have checked that with the present configuration all of them are satisfied up
to $\mathcal{O}(\alpha'^{2})$ except for the $uu$ one, which gives the
following equation for $f_{+}$:

\begin{equation}
\frac{1}{\rho^{3}}\frac{d}{d\rho}\left(\rho^{3} \frac{df_{+}}{d\rho}\right)
=
-\frac{128 \mathcal{Q}_{+}\mathcal{Q}_{-}
\left(\mathcal{Q}_{0}^{2}+\mathcal{Q}_{-}^{2}
+3\mathcal{Q}_{0}\rho^{2}+3\mathcal{Q}_{-}\rho^{2}
+3\rho^{4}+\mathcal{Q}_{0}\mathcal{Q}_{-}\right)}{(\mathcal{Q}_{0}
+\rho^{2})^{3}(\mathcal{Q}_{-}+\rho^{2})^{3}}\, .
\end{equation}

\noindent
where the right-hand side is proportional to $\hat{T}^{(2)}{}_{uu}$.  This
equation is solved by

\begin{equation}
\label{eq:f+}
f_{+}(\rho)
=
-\frac{16\mathcal{Q}_{+}\mathcal{Q}_{-}}{\rho^{6}
\mathcal{Z}^{(0)}_{0}
\mathcal{Z}^{(0)}_{-}}\, ,
\end{equation}

\noindent
up to an arbitrary harmonic function $c_{+}/\rho^{2}$ to be chosen according
to our prescription.

In the $\rho\to 0$ limit the above $f_{+}(\rho)$ diverges as
$-16\mathcal{Q}_{+}\mathcal{Q}_{0}^{-1}/\rho^{2}$ if $\mathcal{Q}_{0}\neq
0$. Then, we choose $c_{+}=+16\mathcal{Q}_{+}\mathcal{Q}_{0}^{-1}$ and we are
left with

\begin{equation}
\label{eq:f+2}
f_{+}(\rho)
=
\frac{16\mathcal{Q}_{+}(\rho^{2}+\mathcal{Q}_{0}+\mathcal{Q}_{-})}
{\mathcal{Q}_{0}(\rho^{2}+\mathcal{Q}_{0})(\rho^{2}+\mathcal{Q}_{-})}\, ,
\end{equation}

\noindent
which has the same structure as $f_{0}(\rho)$ without the corrections
associated to the instanton.\footnote{Up to a factor of 2 it is identical to
  it if we set $\mathcal{Q}_{-}=\mathcal{Q}_{0}$.} Thus, since in this case
there is no contribution to the gauge fields that could cancel this $\alpha'$
correction, the ``near-horizon'' charge and the total, asymptotic charge
associated to $\mathcal{Z}_{+}$ (total momentum) are different and we are
faced with the problem of deciding which of them represents the momentum of
the string. We will discuss this issue in Section~\ref{sec-chargecorrections}.

If $\mathcal{Q}_{0}=0$, $f_{+}\sim 1/\rho^{4}$ when $\rho\to 0$ and there is
no need to shift it by a harmonic function.

Finally, one can check that the dilaton equation is satisfied up to
$\mathcal{O}(\alpha'^{2})$ terms. 

Summarizing the results of this section, we have constructed a solution of the
Heterotic String effective action to first order in $\alpha'$ of the form
given in Eq.~(\ref{10dmetric}) with the $\mathcal{Z}$ functions given, for
$\mathcal{Q}_{0}\neq 0$ by

\begin{eqnarray}
\label{eq:Zs}
\mathcal{Z}_{0}
& = &
\mathcal{Z}^{(0)}_{0}
+
8\alpha' 
\left[
\frac{\rho^{2}+2\kappa^{2}}{(\rho^{2}+\kappa^{2})^{2}}
-
\frac{\rho^{2}+2\mathcal{Q}_{0}}{(\rho^{2}+\mathcal{Q}_{0})^{2}}
\right]
+\mathcal{O}(\alpha'^{2})\, , 
\\
& & \nonumber \\
\mathcal{Z}_{-}
& = &
\mathcal{Z}^{(0)}_{-}
+
\mathcal{O}(\alpha'^{2})\, ,
\\
& & \nonumber \\
\mathcal{Z}_{+}
& = &
\mathcal{Z}_{+}^{(0)}
+16 \alpha'\frac{\mathcal{Q}_{+}(\rho^{2}+\mathcal{Q}_{0}+\mathcal{Q}_{-})}
{\mathcal{Q}_{0}(\rho^{2}+\mathcal{Q}_{0})(\rho^{2}+\mathcal{Q}_{-})}
+\mathcal{O}(\alpha'^{2})\, ,
\end{eqnarray}

\noindent
and for $\mathcal{Q}_{0}=0$ by 

\begin{eqnarray}
\label{eq:ZsQ0=0}
\mathcal{Z}_{0}
& = &
1
+
8\alpha' 
\frac{\rho^{2}+2\kappa^{2}}{(\rho^{2}+\kappa^{2})^{2}}
+\mathcal{O}(\alpha'^{2})\, , 
\\
& & \nonumber \\
\mathcal{Z}_{-}
& = &
\mathcal{Z}^{(0)}_{-}
+\mathcal{O}(\alpha'^{2})\, ,
\\
& & \nonumber \\
\mathcal{Z}_{+}
& = &
\mathcal{Z}^{(0)}_{+}
-16\alpha'\frac{\mathcal{Q}_{+}\mathcal{Q}_{-}}{\rho^{4}
(\rho^{2}+\mathcal{Q}_{-})}
+\mathcal{O}(\alpha'^{2})\, ,
\end{eqnarray}

\noindent
where $\mathcal{Z}^{(0)}_{0},\mathcal{Z}^{(0)}_{\pm}$ are the pieces of the
functions $\mathcal{Z}^{(0)}_{0},\mathcal{Z}^{(0)}_{\pm}$ of zeroth order in
$\alpha'$, namely the harmonic functions in $\mathbb{E}^{4}$

\begin{equation}
\label{eq:harmonicpieces}
\mathcal{Z}^{(0)}_{0,+,-}
=
1+\frac{\mathcal{Q}_{0,+,-}}{\rho^{2}}\, .
\end{equation}

We would like to stress at this point that the $\mathcal{O}(\alpha'^{2})$
terms that we have ignored in the equations of motion derived from the action
Eq.~(\ref{heterotic}) are proportional to products of the Chern-Simons 3-forms
that occur in in $H$. We will discuss in detail in Section~\ref{sec-validity}
when it is justified to disregard these terms as well as the rest of the terms
of higher-order in $\alpha'$ and in the string coupling constant that enter in
the Heterotic Superstring Effective action so, rather than just a solution to
the first-order equations of motion of the Heterotic Superstring Effective
action, we can consider that this is a first-order solution of the full
effective action with second-order corrections in $\alpha'$ and one- and
higher-loop corrections which are negligible when compared with the above
solution.

Let us close this section by commenting the relation of these solutions to the ones 
described in~\cite{Cano:2017nwo}, which were also argued to be exact solutions at
first order in $\alpha'$. Those can be obtained from Eqs.~(\ref{eq:Zs}) by removing the 1's
from the harmonic functions $\mathcal{Z}^{(0)}_{0,+,-}$, but this has the effect of 
removing as well all the $\alpha'$ corrections except the one due to the $SU(2)$ instanton.
The reason is that at zeroth order in $\alpha'$ such solutions are just 
$\mathrm{AdS}_{3}\times\mathrm{S}^{3}\times \mathrm{T}^{4}$, for which $\hat{R}_{(-)}$
vanishes identically~\cite{Prester:2008iu} --- see also Section~\ref{sec-entropy}. 
Hence, only corrections coming from the Yang-Mills fields appear in that case.

\section{The $\alpha'$-corrected charges}
\label{sec-chargecorrections}

Before doing any explicit calculation, it is good to have a more qualitative
discussion on the meaning of the charges that we are going to calculate.

As we have discussed in the previous section, the $\alpha'$ corrections
introduce delocalized terms in the the fields which, generically, give
contributions to the total charges of the fields computed at spatial infinity.
The term in $\mathcal{Z}_{0}$ associated to the presence of the BPST instanton
(let us ignore the second one due to the curvature of the torsionful spin
connection) contributes to the total charge at infinity of the NS 6-form
$\tilde{B}$ dual to the Kalb-Ramond 2-form $B$ and its contribution, which can
be explained in terms of a \textit{gauge 5-brane}
\cite{Strominger:1990et,Cano:2017sqy} is equivalent to that of 8 S5-branes:

\begin{equation}
\mathcal{Z}_{0}
\stackrel{\rho\to\infty}{\sim} 
1+(\mathcal{Q}_{0}+8\alpha')/\rho^{2}+\cdots
\end{equation}

\noindent
If we are interested in finding how many S5-branes there are in the background
this contribution to the total charge must be taken into account and we could
say that $\mathcal{Q}_{0}=N_{\rm S5}\alpha'$. Alternatively, one can look at
the ``near-horizon'' charge which will be determined by the coefficient of the
$1/\rho^{2}$ pole in the $\rho\to 0$ limit. By convention, this is always the
coefficient in $\mathcal{Z}^{(0)}$, $\mathcal{Q}_{0}$:

\begin{equation}
\label{eq:nearhorizonchargeQ0}
\mathcal{Z}_{0}
\stackrel{\rho\to 0}{\sim} 
\mathcal{Q}_{0}/\rho^{2}+\cdots
\end{equation}

Now, let us take into account the  second term in
$f_{0}$ associated to the $R_{(-)}\wedge R_{(-)}$ term. This term contributes to
the total charge at infinity as -8 S5-branes:

\begin{equation}
\mathcal{Z}_{0}
\stackrel{\rho\to\infty}{\sim} 
1+(\mathcal{Q}_{0}+8\alpha'-8\alpha')/\rho^{2}+\cdots
\end{equation}

\noindent
and, therefore, the total charge and the ``near-horizon charge'' which is
always given by Eq.~(\ref{eq:nearhorizonchargeQ0}) are both equal to
$\mathcal{Q}_{0}$ in this case. We do not know of any delocalized extended
object to which this negative contribution to the charge can be attributed to
but the net effect is that we do not need to worry about the different
contributions to this charge. 

Now we can try to use more rigorous definitions (which, of course, will give
the same result). 

In order to compute 5-brane charge we need to use NS 6-form $\tilde{B}$ dual
to the Kalb-Ramond 2-form $B$. The equation of motion of the latter can be
written in the form

\begin{equation}
d\left(e^{-2\phi}\star {H} +\mathcal{O}(\alpha')\right)=0\, ,  
\end{equation}

\noindent
where, according to the discussion in Section~\ref{sec-heteroticalpha} the
$+\mathcal{O}(\alpha')$ terms are related to the zeroth-order equations of
motion. Locally, the equation of motion is solved by 

\begin{equation}
\label{eq:2to6duality}
e^{-2\phi}\star H +\mathcal{O}(\alpha') \equiv d\tilde{B}\, ,
\,\,\,\,
\Rightarrow
\,\,\,\,
H = e^{2\phi}\star\tilde{H}\, , 
\,\,\,\,
\mbox{with}
\,\,\,\,
\tilde{H} \equiv d\tilde{B}+\mathcal{O}(\alpha')\, .
\end{equation}

\noindent
The 6-form equation of motion can be obtained from the Bianchi identity of $H$
Eq.~(\ref{eq:BianchiH})

\begin{equation}
\label{eq:eomB6}
d(e^{2\phi}\star\tilde{H})
-
2\alpha'  \left({F}^{A}\wedge {F}^{A} 
+{R}_{(-)}{}^{{a}}{}_{{b}}\wedge 
{R}_{(-)}{}^{{b}}{}_{{a}}\right)
=0\, ,
\end{equation}

\noindent
and, if we couple the system to $N_{S5}$ solitonic 5-branes lying in the
directions $\tfrac{1}{2}(u+v),y^{1},\cdots,y^{4}$, it takes the
form\footnote{Here we have used the normalization of the Heterotic Superstring
  effective action in Eq.~(\ref{heterotic}), the normalization of the
  Wess-Zumino term of the S5-branes $N_{S5}T_{S5}\, g_{s}^{2}\int
  \phi_{*}\tilde{B}$ and the values of the 10-dimensional Newton constant
  Eq.~(\ref{eq:d10newtonconstant}) and the S5-brane tension in terms of the
  string length $\ell_{s}^{2}=\alpha'$ and the string coupling constant
  $g_{s}$, which are given by
\begin{equation}
T_{S5}= \frac{1}{(2\pi\ell_{s})^{5}\ell_{s} g_{s}^{2}}\, .
\end{equation}
} 

\begin{equation}
\label{eq:S5branecharge}
d(\star e^{2\phi}\tilde{H}) 
-2\alpha'  \left({F}^{A}\wedge {F}^{A} 
+{R}_{(-)}{}^{{a}}{}_{{b}}\wedge 
{R}_{(-)}{}^{{b}}{}_{{a}}\right)
=
4\pi^{2}\alpha' N_{S5} \star_{(4)}\delta^{(4)}(\rho)\, .
\end{equation}

\noindent
This identity means that the expression in the left-hand side is sensitive to
the $1/\rho^{2}$ poles in the $\rho\to 0$ limit and, therefore, the
``near-horizon charge'' $\mathcal{Q}_{0}$ essentially counts the number of
S5-branes in the background, as we explained before. This can be checked
explicitly by using the form Eq.~(\ref{eq:BIanchidigested2}) for the Bianchi
identity in the above expression, and the conclusion is that\footnote{When
  $\kappa=0$ there is an additional contribution to the pole equivalent to 8
  S5-branes that, as we said before, we will simply absorb into a redefinition
  of $\mathcal{Q}_{0}$ so the above identification will always hold.}

\begin{equation}
\label{eq:Q0charge}
\mathcal{Q}_{0}
=
N_{S5}\ \alpha'\, .
\end{equation}

If, instead of the number of S5-branes, we wanted to calculate the total
5-brane charge at infinity, we should move the $\alpha'$ terms to the
right-hand side

\begin{equation}
\label{eq:5branecharge}
d(\star e^{2\phi}\tilde{H}) 
=
4\pi^{2}\alpha' N_{S5} \star_{(4)}\delta^{(4)}(\rho)
-2\alpha'  \left({F}^{A}\wedge {F}^{A} 
+{R}_{(-)}{}^{{a}}{}_{{b}}\wedge 
{R}_{(-)}{}^{{b}}{}_{{a}}\right)
\, ,
\end{equation}

\noindent
and integrate over the 4-dimensional transverse space to the 5-branes. The
total charge would be $(N_{\rm S5}+8N_{\rm G5}-8N_{\rm U5})\alpha'$ where
$N_{\rm G5}$ is the number of gauge 5-branes and is equal to the instanton
number of the gauge field and $N_{\rm U5}$ is the number of ``unknown
5-branes'' associated to the torsionful spin connection $\Omega_{(-)}$ and is
equal to its instanton number too. In our solution $N_{\rm U5}=1$, and, is we
include the $\mathrm{SU}(2)$ instanton, $N_{\rm G5}=1$. Then, the total
5-brane charge is, again, given by Eq.~(\ref{eq:Q0charge}).

Finally, observe that, in the end, the $\alpha'$ terms in the Bianchi identity
are simply those in Eq.~(\ref{eq:equationforf0}) and, as discussed above, only
the $1/\rho^{2}$ poles in $f_{0}$ give contributions to the $\delta$-function.

Let us now move to the fundamental string charge (winding number), described
by $\mathcal{Z}_{-}$ which is not affected by $\alpha'$ corrections. Repeating
the discussion at the beginning of this section we would conclude that the
``near-horizon charge'' and the total charge at infinity should both be equal
to $\mathcal{Q}_{-}$ which, in its turn, should be proportional to the winding
number.

In fact, following Ref.~\cite{Cano:2017qrq} if we have $N_{F1}$ fundamental
strings lying in the direction $\tfrac{1}{2}(u-v)$ we have

\begin{equation}
T_{F1}N_{F1} 
= 
\frac{g_{s}^{2}}{16\pi G_{N}^{(10)}}\int_{V^{8}} 
d\left(\star e^{-2{\phi}}
H +\mathcal{O}(\alpha')\right)\, , 
\hspace{.8cm}
\mbox{where}
\hspace{.8cm}
T_{F1} = \frac{1}{2\pi\alpha'}\, ,
\end{equation}

\noindent
where $\mathcal{O}(\alpha')$ are terms associated to the zeroth-order
equations of motion, as we have discussed, and where $V^{8}$ is the space
transverse to worldsheet parametrized by $u$ and $v$, whose boundary is the
product $\mathrm{T}^{4}\times S^{3}_{\infty}$. The $\mathcal{O}(\alpha')$
terms do not contribute to this integral for the same reason they do not
introduce $\alpha'$-corrections in $\mathcal{Z}_{-}$, which remains a harmonic
function whose pole is the sole contribution to the above integral (see
Eq.~(\ref{eq:equationforf-})). Therefore, using Stokes' theorem and the value
of volume of the $\mathrm{T}^{4}$, $(2\pi \ell_{S})^{4}$, we get again

\begin{equation}
\label{eq:Q-charge}
\mathcal{Q}_{-} = \ell_{s}^{2}g_{s}^{2}N_{F1}\, .  
\end{equation}

Following the same reasoning, the strings' momentum can be found by just
looking at the coefficient of the $1/\rho^{2}$ pole in $\mathcal{Z}_{+}$ which
we have denoted, according to the general convention, by $\mathcal{Q}_{+}$:

\begin{equation}
\label{eq:Q+charge}
\mathcal{Q}_{+} 
=
\frac{g_{s}^{2}\ell_{s}^{4}}{R_{z}^{2}}N_{W}\, .
\end{equation}

However, in this case, the total momentum at infinity is different because
there is a first-order in $\alpha'$ delocalized contribution in $f_{+}(\rho)$
which is not cancelled by the Yang-Mills field's contribution:

\begin{equation}
\label{eq:asumpZ+}
\mathcal{Z}_{+}
\stackrel{\rho\to\infty}{\sim} 
1+\mathcal{Q}_{+}\left(1+16\alpha'/\mathcal{Q}_{0}\right)/\rho^{2}+\cdots
\end{equation}

If $\mathcal{Q}_{0}$ is small, the difference between the string's momentum,
which we have argued should be measure in the near-horizon limit, and the
total, asymptotic momentum, which is assumed to be the momentum of the string
in some of the literature, can be large and with important physical
consequences, as we are going to see in Section~\ref{sec-massless}.

\section{$\alpha'$-corrected T-duality}
\label{sec-T-duality}

In Ref.~\cite{Cano:2017qrq} we arrived to the relation between
$\mathcal{Q}_{+}$ and $N_{\rm W}$ Eq.~(\ref{eq:Q+charge}) via a T-duality
transformation of the solution, which is commonly understood to interchange
momentum and winding of a fundamental string wrapped on a circle. We can call

\begin{equation}
\label{eq:microTdualityrules}
N_{\rm F1}' = N_{\rm W}\, ,
\hspace{1cm}
N_{\rm W}' = N_{\rm F1}\, ,  
\end{equation}

\noindent
the ``microscopic T-duality rules''.  However, these microscopic T-duality
rules come form the study of the Heterotic String spectrum on
$\mathbb{M}^{1,8}\times S^{1}$, in absence of any other background field, but
the system under consideration contains a non-perturbative S5-brane wrapped
around the T-duality direction and it is conceivable that the string spectrum
and the microscopic T-duality rules Eq.~(\ref{eq:microTdualityrules}), which
should be supplemented by 

\begin{equation}
\label{eq:microTdualityrules2}
g_{s}'= g_{s}\ell_{s}/R_{x}\, ,
\hspace{1cm}
R_{x}' = \ell_{s}^{2}/R_{x}\, ,  
\end{equation}

\noindent
suffer $\alpha'$ corrections.

In order to clarify this point we are going to perform a T-duality
transformation of the solution in the direction of propagation of the wave
$x\equiv \tfrac{1}{2}(u-v)$ using the $\alpha'$-corrected Buscher T-duality
rules of Ref.~\cite{Bergshoeff:1995cg} (for $\mu ,\nu\neq \underline{x}$):

\begin{equation}
\label{eq:buscheralphaprime}
\begin{array}{rclrcl}
g'_{\mu\nu} & = &
g_{\mu\nu}+\left[ g_{\underline{x}\underline{x}}
G_{\underline{x}\mu}G_{\underline{x}\nu}
-2G_{\underline{x}\underline{x}}G_{\underline{x}(\mu}
g_{\nu)\underline{x}}\right]/G_{\underline{x}\underline{x}}^{2}\, ,
\hspace{-4cm}
\\
& & \\
B'_{\mu\nu} & = &
B_{\mu\nu}-G_{\underline{x}[\mu}
G_{\nu]\underline{x}}/G_{\underline{x}\underline{x}}\, ,
\\
& & \\
g'_{\underline{x}\mu} 
& = & 
-g_{\underline{x}\mu}/G_{\underline{x}\underline{x}}
+g_{\underline{x}\underline{x}}G_{\underline{x}\mu}
/G_{\underline{x}\underline{x}}^{2}\, , 
\hspace{1cm}
&
B'_{\underline{x}\mu} 
& = &
-B_{\underline{x}\mu}/G_{\underline{x}\underline{x}}
-G_{\underline{x}\mu}/G_{\underline{x}\underline{x}}\, ,
\\
& &
\\
g'_{\underline{x}\underline{x}} 
& = &
g_{\underline{x}\underline{x}}/G_{\underline{x}\underline{x}}^{2}\, , 
&
e^{-2\phi'}
& = & 
e^{-2\phi}|G_{\underline{x}\underline{x}}|\, ,
\\
& &
\\
A'^{A}_{\underline{x}} 
& = & 
-A^{A}_{\underline{x}}/G_{\underline{x}\underline{x}}\, , 
&
A'^{A}_{\mu} 
& = &
A^{A}_{\mu} -A^{A}_{\underline{x}}G_{\underline{x}\mu}/G_{\underline{x}\underline{x}}\, ,
\end{array}
\end{equation}

\noindent 
where $G_{\mu\nu}$ is defined by 

\begin{equation}
G_{\mu\nu}
\equiv
g_{\mu\nu}
-B_{\mu\nu}-2\alpha'
\left\{
A^{A}_{\mu}A^{A}{}_{\nu}
+
\Omega_{(-)\, \mu}{}^{a}{}_{b}
{\Omega}_{(-)\, \nu}{}^{b}{}_{a}
\right\}\, .
\end{equation}

Notice that these $\alpha'$-corrected T-duality transformations are only
well-defined if $G_{\underline{x}\underline{x}} \neq 0$. This corresponds to
the first-order deformation of the non-vanishing radii condition at
zeroth-order, which is $g_{\underline{x}\underline{x}} \neq 0$. This issue
becomes relevant for the exotic solutions presented in
Section~\ref{sec-massless}, for which it is not possible to apply the
transformation.

We only need the components
$G_{\underline{x}\underline{x}},G_{\mu\underline{x}},G_{\underline{x}\mu}$. Taking
into account that, in terms of the coordinates $t,x,x^{m},y^{i}$, the metric
and the Kalb-Ramond 2-form (given in Eq.~(\ref{eq:B})) take the
form\footnote{Observe that here it is not possible to shift away the harmonic
  $-\frac{16\mathcal{Q}_{+}\mathcal{Q}_{-}}{\rho^{2}}$ pole. Its presence here
  is the root of the microscopic T-duality rules that we are going to obtain.}

\begin{eqnarray}
ds^{2}
& = &
\frac{(2-\mathcal{Z}_{+})}{\mathcal{Z}_{-}}dt^{2}  
-\frac{(2+\mathcal{Z}_{+})}{\mathcal{Z}_{-}}dx^{2}
-2\frac{\mathcal{Z}_{+}}{\mathcal{Z}_{-}}dtdx
-\mathcal{Z}_{0}dx^{m}dx^{m} -dy^{i}dy^{i}\, ,  
\\
& & \nonumber \\
B
& = &
-\frac{2}{\mathcal{Z}_{-}}dt\wedge dx 
+\tfrac{1}{4}\mathcal{Q}_{0}\cos{\theta}d\varphi\wedge d\psi\, ,  
\end{eqnarray}

\noindent
that $A^{A}_{\underline{x}}=0$ and 

\begin{eqnarray}
\Omega_{(-)\underline{x}}{}^{a}{}_{b}\Omega_{(-)\underline{x}}{}^{b}{}_{a}
& = &
\Omega_{(-)t}{}^{a}{}_{b}\Omega_{(-)\underline{x}}{}^{b}{}_{a}
= 
\Omega_{(-)\underline{x}}{}^{a}{}_{b}\Omega_{(-)t}{}^{b}{}_{a}
= 
2\mathcal{Z}_{0}^{-1}\mathcal{Z}^{-2}_{-}
\partial_{m}\mathcal{Z}_{+}\partial_{m}\mathcal{Z}_{-}
\nonumber \\
& & \nonumber \\
& = &
-\frac{1}{2\mathcal{Z}_{-}}
\left(f_{+}(\rho)-\frac{16\mathcal{Q}_{+}\mathcal{Q}_{-}}{\rho^{2}} \right)
\, ,
\end{eqnarray}

\noindent
the only non-vanishing components of $G_{\mu\nu}$ we are interested in are
given by

\begin{eqnarray}
G_{\underline{x}\underline{x}}
& = & 
-\mathcal{Z}_{-}^{-1} 
\left[
(2+\mathcal{Z}_{+})
-\alpha'
\left(f_{+}(\rho)-\frac{16\mathcal{Q}_{+}}{\mathcal{Q}_{0}\rho^{2}} \right)
\right]
\nonumber \\
& & \nonumber \\
& = &
-\mathcal{Z}_{-}^{-1}
\left[
2+\mathcal{Z}^{(0)}_{+}
+\frac{16\mathcal{Q}_{+}}{\mathcal{Q}_{0}\rho^{2}}
\right]\, ,
\\
& & \nonumber \\
G_{t\underline{x}}
& = & 
\frac{(2-\mathcal{Z}_{+})}{\mathcal{Z}_{-}} 
-2\alpha'\mathcal{Z}_{0}^{-1}\mathcal{Z}^{-2}_{-}
\partial_{m}\mathcal{Z}_{+}\partial_{m}\mathcal{Z}_{-}
\nonumber  \\
& & \nonumber \\
& = &
\mathcal{Z}_{-}^{-1}
\left[2
-\left(
\mathcal{Z}^{(0)}_{+}
+\frac{16\mathcal{Q}_{+}}{\mathcal{Q}_{0}\rho^{2}}\right)
\right]\, ,
\\
& & \nonumber \\
G_{\underline{x}t}  
& = & 
G_{\underline{x}\underline{x}}\, ,
\end{eqnarray}

\noindent
where $f_{+}$ is the function given in Eq.~(\ref{eq:f+2}) 
and $\mathcal{Z}^{(0)}_{+}$ is the piece of  $\mathcal{Z}_{+}$ of zeroth order
in $\alpha'$ defined in Eq.~(\ref{eq:harmonicpieces}).

Observe that $\mathcal{Z}_{+}^{(0)}$ always occurs in the combination

\begin{equation}
\hat{\mathcal{Z}}_{+}^{(0)}  
=
1+\frac{\hat{\mathcal{Q}}_{+}}{\rho^{2}}\, ,
\hspace{1cm}
\hat{\mathcal{Q}}_{+} 
\equiv 
\mathcal{Q}_{+}\left(1+16\alpha'/\mathcal{Q}_{0}\right)\, ,
\end{equation}

\noindent
where, in view of Eq.~(\ref{eq:asumpZ+}) $\hat{\mathcal{Q}}_{+}$ is the total,
asymptotic, momentum.

Applying straightforwardly the above T-duality rules gives the following
solution

\begin{equation}
\label{10dmetricprime}
\begin{array}{rcl}
d{s}'^{2}
& = &
{\displaystyle
\frac{2}{(2+\hat{\mathcal{Z}}_{+}^{(0)})}d\tilde{u}
\left[d\tilde{v} 
- \tfrac{1}{2}(\mathcal{Z}_{-}+\alpha'f'_{-})d\tilde{u}  \right]
-\mathcal{Z}_{0}dx^{m}dx^{m} -dy^{i}dy^{i}\, ,  
}
\\
& & \\
B'
& = &
{\displaystyle
\frac{1}{(2+\hat{\mathcal{Z}}_{+}^{(0)})}d\tilde{u}\wedge d\tilde{v}
+\tfrac{1}{4}\mathcal{Q}_{0}\cos{\theta}d\varphi\wedge d\psi\, ,  
}
\\
& & \\
{A}'^{A}
& = & 
A^{A}\, ,
\\
& & \\
e^{-2{\phi}'}
& = &
{\displaystyle
e^{-2{\phi}_{\infty}}\frac{(2+\hat{\mathcal{Z}}^{(0)}_{+})}{\mathcal{Z}_{0}}\, ,
}
\end{array}
\end{equation}

\noindent
where

\begin{equation}
f'_{-}(\rho)
\equiv
-\frac{16\mathcal{Q}_{+}\mathcal{Q}_{-}}{\rho^{6}
\mathcal{Z}^{(0)}_{0}
(2+\hat{\mathcal{Z}}^{(0)}_{+})}\, ,  
\end{equation}

\noindent
and where we have defined the light-cone coordinates

\begin{equation}
\tilde{v}\equiv 2t\, ,
\hspace{1cm}
\tilde{u} \equiv x'\, .  
\end{equation}

Observe that, at this order in $\alpha'$, we can replace $\mathcal{Q}_{+}$ by
$\hat{\mathcal{Q}}_{+}$ in $f_{-}'$:

\begin{equation}
f'_{-}(\rho)
\equiv
-\frac{16\hat{\mathcal{Q}}_{+}\mathcal{Q}_{-}}{\rho^{6}
\mathcal{Z}^{(0)}_{0}
(2+\hat{\mathcal{Z}}^{(0)}_{+})}\, ,  
\end{equation}

\noindent
and, then, rewrite the combination

\begin{equation}
\mathcal{Z}_{-}+\alpha'f_{-}' 
=
1
+\frac{\mathcal{Q}_{-}(1-16\alpha'/\mathcal{Q}_{0})}{\rho^{2}}
+16\alpha'\frac{\mathcal{Q}_{-}(3\rho^{2}+\mathcal{Q}_{0}+3\hat{\mathcal{Q}}_{+})}{(\rho^{2}+\mathcal{Q}_{0})(3\rho^{2}+\hat{\mathcal{Q}}_{+})}
+
\mathcal{O}(\alpha'^{2})\, ,
\end{equation}

\noindent
or

\begin{equation}
\mathcal{Z}_{-}+\alpha'f_{-}' 
=
\hat{\mathcal{Z}}_{-}
+16\alpha'\frac{\hat{\mathcal{Q}}_{-}(3\rho^{2}+\mathcal{Q}_{0}+3\hat{\mathcal{Q}}_{+})}{(\rho^{2}+\mathcal{Q}_{0})(3\rho^{2}+\hat{\mathcal{Q}}_{+})}
+
\mathcal{O}(\alpha'^{2})\, ,
\end{equation}

\noindent
where we have defined 

\begin{equation}
\hat{\mathcal{Z}}_{-}  
\equiv
1
+\frac{\hat{\mathcal{Q}}_{-}}{\rho^{2}}\, ,
\hspace{1cm}
\hat{\mathcal{Q}}_{-}
\equiv
\mathcal{Q}_{-}(1-16\alpha'/\mathcal{Q}_{0})\, .
\end{equation}

Thus, the T-dual configuration, including the first-order
$\alpha'$-corrections, can be obtained by replacing everywhere in the original
solution

\begin{equation}
\begin{array}{rcl}
\mathcal{Z}^{(0)\prime}_{-}
& = &
2+\hat{\mathcal{Z}}_{+}^{(0)}\, ,
\\
& & \\
\mathcal{Z}^{(0)\prime}_{+}
& = &
\hat{\mathcal{Z}}_{-}^{(0)}\, .
\end{array}
\end{equation}

Since the constant part of the function $\mathcal{Z}_{+}$ in the original
configuration can be shifted via coordinate transformations $v\rightarrow a u$
for any constant $a$,\footnote{It can also be eliminated by T-dualizing in a
  slightly different direction \cite{Chimento:2018kop}.} we conclude that the net
effect of the T-duality transformation at the level of near-horizon charges is

\begin{equation}
\label{eq:chargetransformations}
\begin{array}{rcl}
\mathcal{Q}_{-}'
& = &
\hat{\mathcal{Q}}_{+}
=
\mathcal{Q}_{+}(1+16\alpha'/\mathcal{Q}_{0})\, ,
\\
& & \\
\mathcal{Q}_{+}'
& = &
\hat{\mathcal{Q}}_{-}
=
\mathcal{Q}_{-}(1-16\alpha'/\mathcal{Q}_{0})\, .
\end{array}
\end{equation}

At first sight, these transformation rules are inconsistent with the relations
between the charges $\mathcal{Q}_{+,-}$ and the winding and momentum numbers
$N_{\rm F1},N_{\rm W}$ in Eqs.~(\ref{eq:Q-charge}) and (\ref{eq:Q+charge}) and
the microscopic T-duality rules Eqs.~(\ref{eq:microTdualityrules}) and
(\ref{eq:microTdualityrules2}), but there are some encouraging signs. For
instance, this transformation is an involution to $\mathcal{O}(\alpha'^{2})$
as long as $16\alpha'/\mathcal{Q}_{0}< 1$:

\begin{equation}
\mathcal{Q}_{\mp}''
=
\mathcal{Q}_{\pm}'(1\pm 16\alpha'/\mathcal{Q}_{0})
=
\mathcal{Q}_{\mp}[1-(16\alpha'/\mathcal{Q}_{0})^{2}]
\sim
\mathcal{Q}_{\mp}+\mathcal{O}(\alpha'^{2})\, .
\end{equation}

Then, using Eqs.~(\ref{eq:Q-charge}) and (\ref{eq:Q+charge}), the
transformations Eqs.~(\ref{eq:chargetransformations}) and the T-duality
transformation of the moduli Eq.~(\ref{eq:microTdualityrules2}), which is
still valid in the $\alpha'$-corrected context,\footnote{They follow from
  Eqs.~(\ref{eq:buscheralphaprime}) by restoring the radius of the $x$
  direction so that $g_{\underline{x}\underline{x}}\sim (R_{z}/\ell_{s})^{2}$
  at infinity.} we arrive at the following microscopic T-duality
transformations that replace Eq.~(\ref{eq:microTdualityrules}) in this
context:

\begin{equation}
\label{eq:microTdualityrules3}
N_{\rm F1}' = N_{\rm W}(1+16/N_{\rm S5})\, ,
\hspace{1cm}
N_{\rm W}' = N_{\rm F1}(1-16/N_{\rm S5})\, ,
\end{equation}

\noindent
and which are involutive to second order in $1/N_{\rm S5}$ if $N_{\rm S5}\gg
16$.

The correctness of these rules cannot be showed using the effective field
theory methods used in this paper. It should be mentioned that, had we adopted
the point of view that the asymptotic
$\hat{\mathcal{Q}}_{+}=g_{s}^{2}\ell_{s}^{4}N_{W}/R_{z}^{2}$, the rules
Eq.~(\ref{eq:microTdualityrules}) would still hold. However, since
$\mathcal{Q}_{+}=\hat{\mathcal{Q}}_{+}(1-16\alpha'/\mathcal{Q}_{0})$, it can
become negative for small values of $N_{\rm S5}$, giving rise to 5-dimensional
black holes with regular horizon and negative or vanishing mass. These
pathological solutions disappear if $N_{\rm S5}\gg 1$ because the first-order
$\alpha'$ corrections become very small. We will discuss in
Section~\ref{sec-validity} if it is necessary to impose this condition or not.

\section{$\alpha'$ corrections to the 5-dimensional non-Abelian 
black hole solution}
\label{sec-bhcorrections}

When we compactify the Heterotic Superstring Effective action to first order
in $\alpha'$ on a $T^{5}$ we get a very complicated action with higher-order
terms in curvatures which is very difficult to work with. The definitions of
some gauge fields are also affected by the presence of the Chern-Simons term
of the torsionful spin connection $\Omega_{(-)}$. However, we can just focus
on the metric and the two scalar fields of the 5-dimensional solution ( the
5-dimensional dilaton field $\phi$ and the Kaluza-Klein scalar of the
$6\rightarrow 5$ compactification, $k$), which are obtained from the
10-dimensional one exactly as in absence of $\alpha'$ corrections and take the
form \cite{Cano:2017qrq}

\begin{equation}
\label{eq:3chargebh}
\begin{array}{rcl}
ds^{2} 
& = & 
f^{2}dt^{2}-f^{-1}(d\rho^{2}+\rho^{2}d\Omega_{(3)}^{2})\, ,
\\
& & \\
e^{2\phi}
& = &
e^{2\phi_{\infty}}{\displaystyle\frac{\mathcal{Z}_{0}}{\mathcal{Z}_{-}}}\, ,
\\
& & \\
k
& = & 
k_{\infty}(f \mathcal{Z}_{+})^{3/4}\, ,
\end{array}
\end{equation}

\noindent 
where $\phi_{\infty}$ and $k_{\infty}$ are the asymptotic values of $\phi$ and
$k$, the metric function $f$ is given by

\begin{equation}
\label{eq:f}
f^{-3}
=
\mathcal{Z}_{0}\, \mathcal{Z}_{+}\, \mathcal{Z}_{-}\, ,
\end{equation}

\noindent
and the $\mathcal{Z}$ functions take the form given in Eqs.~(\ref{eq:Zs}) and
(\ref{eq:ZsQ0=0}).





Observe that the $\alpha'$ corrections of $\mathcal{Z}_{0}$ cancel identically
in the $\rho\to \infty$ limit, unless $\mathcal{Q}_{0}=0$, in which case only
the term associated to the Yang-Mills field contributes. The value of its
contribution in that limit is independent of the value of $\kappa$ but,
according to the previous discussions, when $\kappa=0$ this contribution must
be understood as that of 8 S5-branes and we will simply absorb them into
$\mathcal{Q}_{0}=0$.

Taking these considerations and conventions into account, and expressing all
the 5-dimensional constants in terms of the 10-dimensional ones using
Eqs.~(\ref{eq:d10newtonconstant}), (\ref{eq:Q0charge}), (\ref{eq:Q-charge}),
(\ref{eq:Q+charge}) the mass of this family of black-hole solutions is given
by

\begin{eqnarray}
\label{eq:mass}
M
& = &
\frac{\pi}{4G_{N}^{(5)}}  
\left[
\mathcal{Q}_{0}
+\mathcal{Q}_{+}(1+16\alpha'/\mathcal{Q}_{0}) 
+\mathcal{Q}_{-} 
\right]
\nonumber \\
& & \nonumber \\
& = &
\frac{R_{z}}{g^{2}_{s}\ell_{s}^{2}}
N_{\rm S5}
+
\frac{R_{z}}{\ell_{s}^{2}}N_{\rm F1}
+
\frac{1}{R_{z}}N_{\rm W}(1+16/N_{\rm S5})\, ,
\,\,\,\,\,
\text{for}
\,\,\,\,\,
\mathcal{Q}_{0}\neq 0\, ,
\\
& & \nonumber \\  
\label{eq:massQ0=0}
M
& = &
\frac{\pi}{4G_{N}^{(5)}}  
\left[
8\alpha'
+\mathcal{Q}_{+}
+\mathcal{Q}_{-} 
\right]
\nonumber \\
& & \nonumber \\
& = &
8\frac{R_{z}}{g^{2}_{s}\ell_{s}^{2}}
+
\frac{R_{z}}{\ell_{s}^{2}}N_{\rm F1}
+
\frac{1}{R_{z}}N_{\rm W}\, ,
\,\,\,\,\,
\text{for}
\,\,\,\,\,
\mathcal{Q}_{0} = 0\, .
\end{eqnarray}

The mass depends on the total, asymptotic charges and, therefore, written in
terms of the numbers of branes (``near-horizon charges''), contains additional
terms from the delocalized fields.

The area of the horizon, which will give the leading contribution to the
entropy, as we will see in Section~\ref{sec-entropy}, is given by

\begin{eqnarray}
\begin{array}{rcl}
\label{eq:area}
A_{\rm H}
& = &
2\pi^{2} 
\sqrt{ 
\mathcal{Q}_{0} \mathcal{Q}_{+}\mathcal{Q}_{-}
}\, ,
\,\,\,\,\,
\text{for}
\,\,\,\,\,
\mathcal{Q}_{0} \neq 0\, ,
\\
& & \nonumber \\
A_{\rm H}
& = &
2\pi^{2} 
\sqrt{ 
-16\alpha'\mathcal{Q}_{+}\mathcal{Q}_{-}
}\, ,
\,\,\,\,\,
\text{for}
\,\,\,\,\,
\mathcal{Q}_{0} = 0\, .
\\
\end{array}
\end{eqnarray}

In the $\mathcal{Q}_{0} = 0$ case one of the two non-vanishing charges has to
be negative for the horizon to exist at all. If $\mathcal{Q}_{-}<0$ then
$\mathcal{Z}_{-}$ will vanish at $\rho^{2}= |\mathcal{Q}_{-}|$. If
$\mathcal{Q}_{-}<0$ the vanishing of $\mathcal{Z}_{+}$ depends on the values
of $\mathcal{Q}_{+}$ and $\mathcal{Q}_{-}$ and we will explores the different
possibilities in Section~\ref{sec-massless} even though the near-horizon
geometry is singular in $d=10$.

In the next section we consider other possible contributions to the entropy.

\section{BH entropy}
\label{sec-entropy}

In order to find the entropy, we would need to compactify the action down to 5
dimensions and use there Wald's entropy formula
\cite{Wald:1993nt,Iyer:1994ys}

\begin{equation}
\label{eq:Waldformula5}
S 
= 
-2\pi\int_{\rm H}d^{3}x\sqrt{|h|}\, \frac{\partial\mathcal{L}_{(5)}}{\partial
  R_{abcd}}\epsilon_{ab}\epsilon_{cd}\, ,
\end{equation}

\noindent
where $h$ is determinant of the 3-dimensional metric induced on the horizon
$ds_{\rm H}^{2}$, $\epsilon_{ab}$ is the binormal to the bifurcation surface,
normalized as $\epsilon_{ab}\epsilon^{ab}=-2$, $\mathcal{L}_{(5)}$ is the
5-dimensional Lagrangian and $R_{abcd}$ is the 5-dimensional Riemann
tensor.\footnote{All the indices in this expression run from $0$ to
  $4$. 10-dimensional indices will be distinguished with hats in this
  section.}

This formula is valid for diff-invariant theories. The 10-dimensional action
Eq.~(\ref{heterotic}) is, by construction, exactly diff-invariant to first
order in $\alpha'$, and so would be the 5-dimensional theory that follows from
the direct compactification to 5 dimensions.  Therefore, Wald's formula can be
applied to it and no terms such as those considered in
Ref.~\cite{Tachikawa:2006sz} need to be added.

Compactifying the $\alpha'$-corrected action is a very involved calculation
that, quite understandably, we would like to avoid carrying out. Thus, we try
a different strategy, directly applying this formula to the 10-dimensional
action.

First of all, we have to identify the part of the 10-dimensional Riemann
curvature that corresponds to the 5-dimensional one.  The decomposition of the
10-dimensional metric in 5-dimensional variables is given by
\cite{Cano:2017qrq}

\begin{equation}
d\hat s^{2} 
= 
e^{\phi-\phi_{\infty}}\left[(k/k_{\infty})^{-2/3}ds^{2}-k^{2}\mathcal{A}^{2}\right]
-dy^{i}dy^{i}\, ,
\end{equation}

\noindent
where $\mathcal{A}$ is the 1-form

\begin{equation}
\mathcal{A} 
\equiv
dz+\frac{k_{\infty}^{1/3}}{\sqrt{12}}(A^{1}+A^{2})\, ,
\end{equation}

\noindent
and $A^{1},A^{2}$ are certain 5-dimensional vector fields.

If we decompose the 10-dimensional flat and curved indices as, respectively,
$\hat{a}=a\, ,z\, , i$ and $\hat{\mu}=\mu\, ,\underline{z}\, ,\underline{i}$,
the, F\"unfbein $e^{a}{}_{\mu}$ is related to the components
$\hat{e}^{a}{}_{\mu}$ of the Zehnbein $\hat{e}^{\hat{a}}{}_{\hat{\mu}}$ by 

\begin{equation}
\hat{e}^{a}{}_{\mu}=e^{(\phi-\phi_{\infty})/2}(k/k_{\infty})^{-1/3}
e^{a}{}_{\mu}\, ,  
\end{equation}

\noindent
so the 5-dimensional Riemann curvature $R_{abcd}$ is related to the
$\hat{R}_{abcd}$ components of the 10-dimensional Riemann curvature
$\hat{R}_{\hat{a}\hat{b}\hat{c}\hat{d}}$ by

\begin{equation}
\hat{R}_{abcd}=e^{-(\phi-\phi_{\infty})}(k/k_{\infty})^{2/3} R_{abcd}+\ldots\, .
\end{equation}

\noindent
Furthermore, the 10-dimensional Riemann curvature enters the curvature tensor
of the torsionful spin connection $\hat{R}_{(-)\,
  \hat{a}\hat{b}\hat{c}\hat{d}}$ in this way

\begin{equation}\label{R-def}
\hat{R}_{(-)\, \hat{a}\hat{b}\hat{c}\hat{d}}
=
\hat{R}_{\hat{a}\hat{b}\hat{c}\hat{d}}
-\hat{\nabla}_{[\hat{a}}\hat{H}_{\hat{b}]\hat{c}\hat{d}}
+\tfrac{1}{2}\hat{H}_{[\hat{a}|\hat{c}\hat{e}}\hat{H}_{|\hat{b}]\hat{d}}{}^{\hat{e}}\, ,
\end{equation}

\noindent
so 

\begin{equation}
\hat{R}_{(-)\, abcd}
=
e^{-(\phi-\phi_{\infty})}(k/k_{\infty})^{2/3} R_{abcd}+\ldots\, .
\end{equation}

Taking into account these relations, Wald's entropy formula
Eq.~(\ref{eq:Waldformula5}) can be rewritten in terms of the 10-dimensional
Lagrangian and the 10-dimensional Riemann tensor for the family of metrics
under consideration as\footnote{The reason why the metric function appears
  explicitly is because it is the optimal way of taking into account the
  rescalings the action goes through in the dimensional reduction. We can
  write, for these metrics,
\begin{equation}
\sqrt{|h|} \mathcal{L}_{(5)} 
= 
\frac{\sqrt{|g|}}{\sqrt{f}} \mathcal{L}_{(5)}
=
\frac{\sqrt{|\hat{g}|}}{\sqrt{f}}\, 
\mathcal{L}_{(10)}\, ,
\end{equation}
because the Lagrangian density is the same in any dimension.
}

\begin{equation}
S
=
-2\pi\int_{\mathrm{H}\times \mathrm{S}^{1}\times \mathrm{T}^{4}} d^{8}\hat{x}
\frac{\sqrt{|\hat{g}|}}{\sqrt{f}}\, 
e^{-(\phi-\phi_{\infty})}(k/k_{\infty})^{2/3}
\frac{\partial\mathcal{L}_{(10)}}{\partial \hat{R}_{abcd}}\epsilon_{ab}\epsilon_{cd}\, ,
\end{equation}

\noindent
where $|\hat{g}|$ is the absolute value of the full 10-dimensional metric and
we we are integrating over the co-dimension 2 surface $\mathrm{H}\times
\mathrm{S}^{1}\times \mathrm{T}^{4}$, and where the binormal $\epsilon_{ab}$
is intrinsically 5-dimensional. In the Vielbein basis, though, $\epsilon_{ab}$
has the same components both in the 5-dimensional and in the 10-dimensional
basis.

Let us apply this formula to the different pieces of the 10-dimensional action
that contain the 10-dimensional Riemann tensor, manifestly, or via
$\hat{R}_{(-)}$. 

Applied to the Riemann-Hilbert term, we have

\begin{eqnarray}
\frac{\partial\mathcal{L}_{(10)}}{\partial \hat{R}_{abcd}}
& = &
\frac{1}{16\pi G_{N}^{(10)}} e^{-2(\hat{\phi}-\hat{\phi}_{\infty})} 
\hat{\eta}^{ac}\hat{\eta}^{bd} \, ,
\\
& & \nonumber \\
\frac{\sqrt{|\hat{g}|}}{\sqrt{f}}
& = & 
\tfrac{1}{8}k_{\infty}e^{+3(\hat{\phi}-\hat{\phi}_{\infty})}(k/k_{\infty})^{-2/3}
(\rho^{6}f^{-3})^{1/2}\sin{\theta}\, ,
\end{eqnarray}

\noindent
where, evidently $\hat{\eta}^{ab}=\eta^{ab}$. Then, taking the
$\rho\rightarrow 0$ limit, substituting in the formula and integrating over
the 5 compact dimensions whose coordinates take values in $[0,2\pi \ell_{s})$,
and over the 3-sphere, and using $k_{\infty}=R_{z}/\ell_{s}$, we get the
zeroth-order contribution to the entropy

\begin{equation}
S^{(0)}=\frac{A_{\rm H}}{4 G_{N}^{(5)}}\, ,
\end{equation}

\noindent
where $A_{\rm H}$ is the area of the horizon, computed in Eq.~(\ref{eq:area}) and
where the 5-dimensional Newton constant is

\begin{equation}
\frac{1}{G_{N}^{(5)}}
=
\frac{(2\pi \ell_{s})^{4}(2\pi R_{z})}{G_{N}^{(10)}}\, .
\end{equation}

Using the result Eq.~(\ref{eq:area}) and the relations between the 5- and
10-dimensional constants, this zeroth-order contribution is, in terms of the
brane numbers

\begin{equation}
\label{eq:entropy0}
S^{(0)}
= 
2\pi
\sqrt{ 
N_{\rm S5}N_{\rm F1}N_{\rm W}
}\, ,
\end{equation}

\noindent
which is the classical, zeroth-order in $\alpha'$ result.

The are two terms that contribute to Wald's formula at first order in
$\alpha'$ through the occurrence of $\hat{R}_{(-)}$: the kinetic term of the
Kalb-Ramond field, whose field strength contains $\hat{R}_{(-)}$ in the
Lorentz-Chern-Simons term, and in the $\hat{T}^{(2)}$ tensor. Let us start
with the latter.

The contribution of the $\hat{T}^{(0)}$ tensor term of the action to Wald's
formula is clearly proportional to $\hat{R}_{(-)}$. However, when evaluated on
$\mathrm{AdS}_{3}\times\mathrm{S}^{3}\times \mathrm{T}^{4}$, $\hat{R}_{(-)}$
vanishes identically \cite{Prester:2008iu}. It is easy to prove this fact
explicitly: the Riemann tensor takes the form

\begin{equation}
\hat{R}_{\hat{a}\hat{b}\hat{c}\hat{d}}
=
\left(
-\frac{2}{L^{2}}\hat g_{\hat{a}[\hat{c}}\hat g_{\hat{d}]\hat{b}}\, ,\,
\frac{2}{L^{2}}\hat g_{\hat{a}[\hat{c}}\hat g_{\hat{d}]\hat{b}}\, ,0\right)
\, ,
\end{equation}

\noindent
in a more or less obvious notation in which each factor corresponds,
respectively, to $\mathrm{AdS}_{3}$, $\mathrm{S}^{3}$ and $\mathrm{T}^{4}$,
and $L$ is the common radius of $\mathrm{AdS}_{3}$ and of the sphere. Only the
indices corresponding to those subspaces are active in each factor, but we
will not introduce new indices to keep the notation as simple as possible.

On the other hand, the 3-form field strength can be put in the form

\begin{equation}
\hat{H}=\frac{2}{L}\left(-d\Pi_{3}+dV_{3}\right)\, ,
\end{equation}

\noindent
where $d\Pi_{3}$ is the volume form of the $\mathrm{AdS}_{3}$ factor (with
unit radius) and $dV_{3}$ the volume form of $\mathrm{S}^{3}$ (of unit radius
too). Then, $\hat{H}$ is covariantly constant, $\hat{\nabla}\hat{H}=0$, and we
can see that, in the same notation,

\begin{equation}
\hat{H}_{[\hat{a}|\hat{c}\hat{e}}\hat{H}_{|\hat{b}]\hat{d}}{}^{\hat{e}}
=
\left(
+\frac{4}{L^{2}}\hat{g}_{\hat{a}[\hat{c}}\hat{g}_{\hat{d}]\hat{b}}\, ,\, 
-\frac{4}{L^{2}}\hat{g}_{\hat{a}[\hat{c}}\hat{g}_{\hat{d}]\hat{b}}\, ,0\right)\, ,
\end{equation}

\noindent
which implies, according to the definition Eq.~(\ref{R-def}) $\hat{R}_{(-)\,
  \hat{a}\hat{b}\hat{c}\hat d}=0$. Since $\mathrm{AdS}_{3}\times
\mathrm{S}^{3}\times \mathrm{T}^{4}$ is the near-horizon of extremal black
holes as the ones we are considering, we conclude that for these extremal
black holes the $\hat{T}^{(0)}$-tensor term in the action does not contribute
to Wald's entropy formula.

Then, the only possible first-order contribution comes from 

\begin{eqnarray}
S^{(1)}
& = &
-2\pi
\int 
d^{8}\hat{x}
\frac{\sqrt{|\hat{g}|}}{\sqrt{f}}\, 
e^{-(\phi-\phi_{\infty})}(k/k_{\infty})^{2/3}
\frac{\partial~}{\partial \hat{R}_{abcd}}
\left\{
\tfrac{1}{2\cdot 3!} \frac{e^{-2(\hat{\phi}-\hat{\phi}_{\infty})}}{16\pi G_{N}^{(10)}}
\hat{H}^{2}
\right\}\epsilon_{ab}\epsilon_{cd}
\nonumber \\
& & \nonumber \\
& = & 
-\frac{1}{48 G_{N}^{(10)}}
\int d^{8}\hat{x}
\frac{\sqrt{|\hat{g}|}}{\sqrt{f}}\, 
e^{-3(\phi-\phi_{\infty})}(k/k_{\infty})^{2/3}
\hat{H}^{\hat{e}\hat{f}\hat{g}}
\frac{\partial \hat{H}_{\hat{e}\hat{f}\hat{g}}}{\partial \hat{R}_{abcd}}
\epsilon_{ab}\epsilon_{cd}
\nonumber \\
& & \nonumber \\
& = & 
\frac{\alpha'}{8 G_{N}^{(10)}}
\int d^{8}\hat{x}
\frac{\sqrt{|\hat{g}|}}{\sqrt{f}}\, 
e^{-3(\phi-\phi_{\infty})}(k/k_{\infty})^{2/3}
\hat{H}^{ab\hat{g}}\hat{\Omega}_{(-)\,
  \hat{g}}{}^{cd}\epsilon_{ab}\epsilon_{cd}\, .
\end{eqnarray}

The binormal has the following components:\footnote{The global sign is
  irrelevant, as it appears twice in the formula.} $\epsilon_{0\sharp}=1$,
where $e^{\sharp}=f^{-1/2}d\rho$ and, therefore,

\begin{equation}
S^{(1)}
=
\frac{\alpha'}{2 G_{N}^{(10)}}
\int d^{8}\hat{x}
\frac{\sqrt{|\hat{g}|}}{\sqrt{f}}\, 
e^{-3(\phi-\phi_{\infty})}(k/k_{\infty})^{2/3}
\hat{H}^{0\sharp\hat{g}}\hat{\Omega}_{(-)\, \hat{g}}{}^{0\sharp}\, .
\end{equation}

\noindent
In Appendix~\ref{sec-connection} we have computed explicitly the components of
$\hat{H}$ (Eq.~(\ref{eq:Hcomponents})) using the Zehnbein basis
Eq.~(\ref{eq:zehnbein}), but this is not the basis related by a simple
rescaling to the F\"unfbein basis in which $\epsilon_{0\sharp}=1$. The
relation is

\begin{equation}
\begin{array}{rcl}
\hat{e}^{0} 
& = & 
{\displaystyle 
\tfrac{1}{2}
\sqrt{\mathcal{Z}_{+}\mathcal{Z}_{-}}\,\hat{e}^{+}
+\frac{1}{\sqrt{\mathcal{Z}_{+}\mathcal{Z}_{-}}}\, \hat{e}^{-} 
\, ,
}
\\
& & \\
\hat{e}^{1} 
& = &
{\displaystyle 
\tfrac{1}{2}
\sqrt{\mathcal{Z}_{+}\mathcal{Z}_{-}}\hat{e}^{+}
-\frac{1}{\sqrt{\mathcal{Z}_{+}\mathcal{Z}_{-}}}\hat{e}^{-} 
\, ,
}
\\
& & \\
\hat{e}^{\sharp}
& = &
{\displaystyle 
\frac{x^{m}}{\rho}\hat{e}^{m}\, ,  
}
\end{array}
\end{equation}

\noindent
and leads to

\begin{eqnarray}
\hat{H}^{0\sharp\hat{g}} 
& = &
\tfrac{1}{2}
\delta^{\hat{g}}{}_{-}
\sqrt{\mathcal{Z}_{+}\mathcal{Z}_{-}}\frac{x^{m}}{\rho}\hat{H}^{+m-}
+\delta^{\hat{g}}{}_{+}
\frac{1}{\sqrt{\mathcal{Z}_{+}\mathcal{Z}_{-}}}\frac{x^{m}}{\rho}\hat{H}^{-m+} 
\, ,
\\
& & \nonumber \\ 
\hat{\Omega}_{(-)\, \hat{g}}{}^{0\sharp}
& = & 
\delta_{\hat{g}}{}^{+}
\left(\tfrac{1}{2}\sqrt{\mathcal{Z}_{+}\mathcal{Z}_{-}}
\frac{x^{m}}{\rho}\hat{\Omega}_{(-)\, +}{}^{+m}
+\frac{1}{\sqrt{\mathcal{Z}_{+}\mathcal{Z}_{-}}}\frac{x^{m}}{\rho}
\hat{\Omega}_{(-)\, +}{}^{-m}\right)\, ,
\\
& & \nonumber \\
\hat{H}^{0\sharp\hat{g}} \hat{\Omega}_{(-)\, \hat{g}}{}^{0\sharp}
& = & 
-\frac{x^{m}}{\rho}\hat{H}_{+-m} 
\frac{x^{n}}{\rho}\left( \tfrac{1}{2}\hat{\Omega}_{(-)\, +-n} 
+\frac{1}{\mathcal{Z}_{+}\mathcal{Z}_{-}}\hat{\Omega}_{(-)\, ++n}
\right)
\nonumber \\
& & \nonumber \\
& = & 
\frac{1}{2\mathcal{Z}_{0}}\partial_{\rho}\log{\mathcal{Z}_{-}}
\left(\partial_{\rho}\log{\mathcal{Z}_{-}}+\partial_{\rho}\log{\mathcal{Z}_{+}} 
\right)\, .
\end{eqnarray}

Observe that, in the near-horizon ( $\rho\to 0$) limit,
$\partial_{\rho}\log{\mathcal{Z}_{-}}
\left(\partial_{\rho}\log{\mathcal{Z}_{-}}+\partial_{\rho}\log{\mathcal{Z}_{+}}
\right)\sim 1/\rho^{2}$, and the above term will only be finite if, in the
same limit, $\mathcal{Z}_{0}\sim 1/\rho^{2}$, \textit{i.e.}~if
$\mathcal{Q}_{0}\neq 0$. Nevertheless, what really matters is the
$\rho\to 0$ limit of the product of this term with $(\rho^{6}f^{-3})^{1/2}$,
and this limit is finite if the separate limits of the two factors are finite
(this is what happens when all the charges are finite) or when
$\mathcal{Q}_{0}=\mathcal{Q}_{+}=0$, a case in which there is no
classical horizon. For small black holes, this contribution will be divergent.

Then, plugging this result into the above expression for $S^{(1)}$ and
evaluating it for $\mathcal{Q}_{0}\neq 0$, we get

\begin{equation}
S^{(1)} = +\frac{8\alpha'}{\mathcal{Q}_{0}}S^{(0)}\, ,
\end{equation}

\noindent
and, to first order in $\alpha'$, and for $\mathcal{Q}_{0}\neq 0$ the
entropy is given by

\begin{equation}
\label{eq:entropy}
\begin{array}{rcl}
S
& = & 
2\pi
\sqrt{ 
N_{\rm S5}N_{\rm F1}N_{\rm W}}
\left(
1+8/N_{\rm S5}
\right)
\\
& & \\
& \sim &
2\pi
\sqrt{ (N_{\rm S5}+16)N_{\rm F1}N_{\rm W}}\, ,
\,\,\,\,\,
\text{for}
\,\,\,\,\,
N_{\rm S5}\gg 16\, .
\end{array}
\end{equation}

\section{Small black holes}
\label{sec-massless}

Another potentially interesting feature of these $\alpha'$-corrected solutions
which has been observed in the literature before,\footnote{See,
  \textit{e.g.}~Refs.~\cite{Dabholkar:2004dq,Alishahiha:2007ap}, the review
  Ref.~\cite{Mohaupt:2000mj} and references therein.} is the emergence of
regular horizons in certain configurations with only two non-vanishing charges
which, in our case, must be $\mathcal{Q}_{+}$ and $\mathcal{Q}_{-}$. For
$\mathcal{Q}_{0}=0$, the area of the horizon is given by the second equation
in (\ref{eq:area}), which we rewrite here for the sake of convenience:

\begin{equation}
\label{eq:areaQ0=0}
A_{\rm H}
= 
2\pi^{2} 
\sqrt{ 
-16\alpha'\mathcal{Q}_{+}\mathcal{Q}_{-}
}\, .  
\end{equation}

\noindent
This expression can be real and finite, and $\mathcal{Z}_{-}>0\,\, \forall\,
\rho$, if $\mathcal{Q}_{+}<0$ and $\mathcal{Q}_{-}>0$. Now we have to study if
there are values of these constants for which $\mathcal{Z}_{+}(\rho)>0\,\,
\forall\, \rho$, making the 5-dimensional metric completely regular. This
function can be written in the form

\begin{equation}
\mathcal{Z}_{+}
 =
1-\frac{|\mathcal{Q}_{+}|}{\rho^{2}}
\left[
1-\frac{16\alpha'\mathcal{Q}_{-}}{\rho^{2}(\rho^{2}+\mathcal{Q}_{-})}
\right]
+\mathcal{O}(\alpha'^{2})\, . 
\end{equation}

\begin{figure}[t]
  \centering
  \includegraphics[height=8cm]{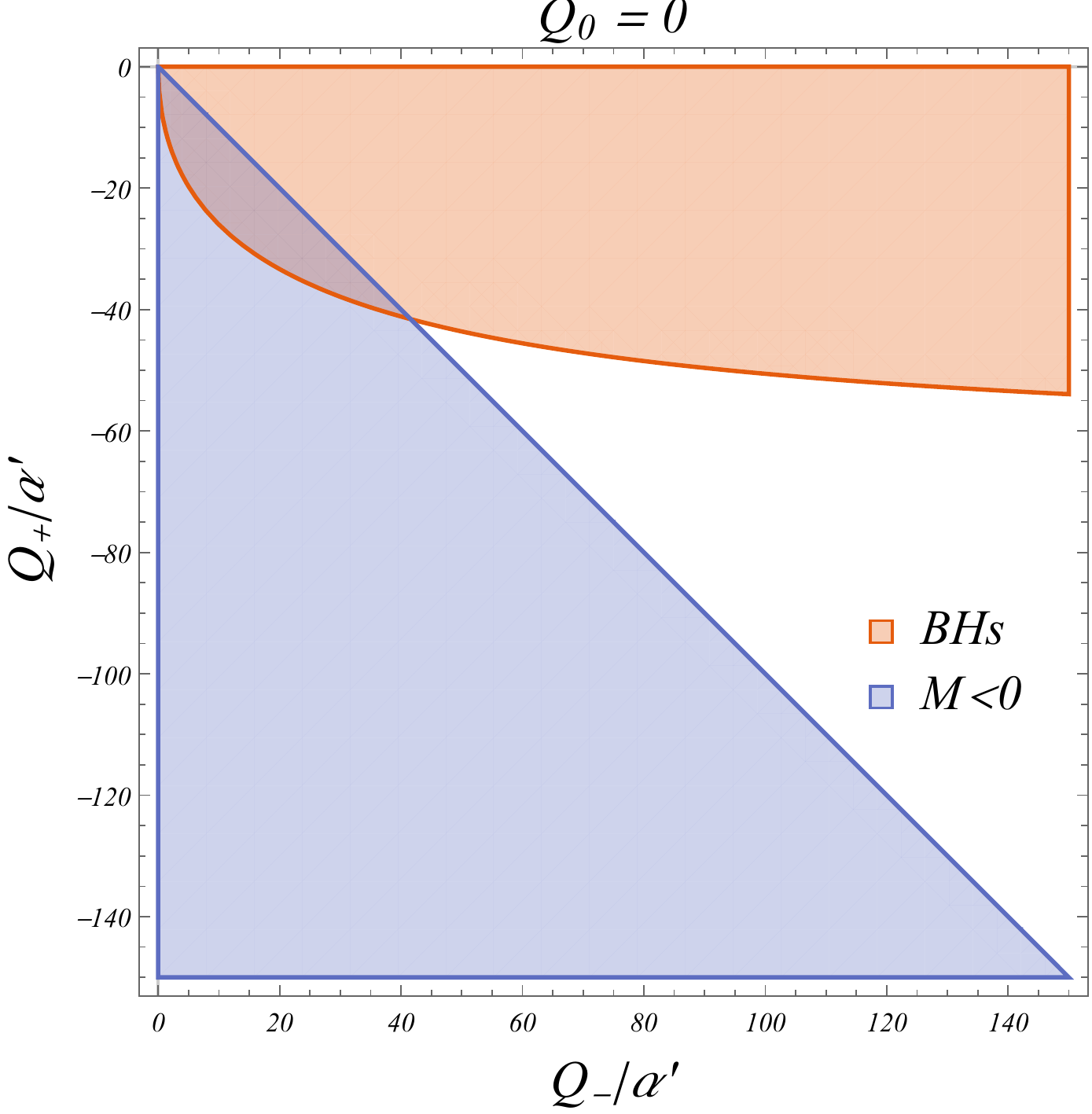}
  \caption{\small Location in $\mathcal{Q}_{+}$-$\mathcal{Q}_{-}$ charge space
    of the small black holes ($\mathcal{Q}_{0}=0$) whose horizon area
    is rendered finite due to the $\alpha'$ corrections in their geometry.}
  \label{fig:Q0}. 
\end{figure}

It is not difficult to see that there are values of $\mathcal{Q}_{+}<$ and
$\mathcal{Q}_{-}>0$ for which the regularity condition is satisfied.  The
orange-shaded region in Figure~\ref{fig:Q0} corresponds to the values of
$\mathcal{Q}_{+},\mathcal{Q}_{-}$ for which the black holes have a regular
horizon due to the $\alpha'$ corrections. The blue-shaded area corresponds to
the small black holes with $|\mathcal{Q}_{+}|\geq \mathcal{Q}_{-}$, which have
negative or vanishing mass. 

For $\mathcal{Q}_{-}\gg \alpha'$ it is possible to see that the condition on
the other charge is $0>\mathcal{Q}_{+}> - 64\alpha'$. Thus, the small black
holes are confined to the region of small $\mathcal{Q}_{+}$.

\section{Range of validity of the solution}
\label{sec-validity}

So far we have studied the solutions ignoring whether they are really good
solutions of the complete Heterotic Superstring effective action to first
order in $\alpha'$ and to zeroth order in the string coupling constant,
everywhere.

Let us start with the possible loop corrections. These will be small if

\begin{equation}
e^{\phi}=e^{\phi_{\infty}}\sqrt{\mathcal{Z}_{0}/\mathcal{Z}_{-}}\, ,
\end{equation}

\noindent
whose vacuum expectation value gives the string coupling constant, is
small. For $\mathcal{Q}_{0}\neq 0$, it is easy to see that, at spatial
infinity, this requires 

\begin{equation}
e^{\phi_{\infty}}=g_{s}\ll 1\, ,
\end{equation}

\noindent
while at the horizon (and also at intermediate values of $\rho$) this requires

\begin{equation}
\label{eq:Q-Q0}
\mathcal{Q}_{-}\lesssim \mathcal{Q}_{0} \, ,
\,\,\,\,\,
\text{or}
\,\,\,\,\,
N_{\rm F1}\gg N_{\rm S5}\, .
\end{equation}

\noindent
For $\mathcal{Q}_{0} = 0$,
the dilaton vanishes at $\rho=0$ and there is no need to impose any more
conditions.

Another important condition that the solution must satisfy is that the radius
of compactification of the 6th dimension, measured in $\ell_{s}$ units by
$|g_{zz}|^{1/2}$ in the 10-dimensional string frame

\begin{equation}\label{gzz}
|g_{zz}|^{1/2}
=
ke^{\tfrac{1}{2}(\phi-\phi_{\infty})}
=
k_{\infty}\sqrt{\left|\frac{\mathcal{Z}_{+}}{\mathcal{Z}_{-}}\right|}\, ,
\end{equation}

\noindent
is much larger than the self-dual\footnote{Self-dual under T-duality.} radius
$\sim \ell_{s}$, at which new massless modes appear in the string spectrum
that invalidate the effective action we have used because they have not been
taken into account in it. At infinity, this condition,
$|g_{zz}|^{1/2}\gg 1$, translates into

\begin{equation}
k_{\infty} \gg 1\, ,
\,\,\,\,\,
\Rightarrow
\,\,\,\,\,
R_{z}\gg \ell_{s}\, .  
\end{equation}

\noindent
If $\mathcal{Q}_{0}=0$, $|g_{zz}|^{1/2}$ diverges in the $\rho\to 0$
limit and, again, no conditions must be imposed on the remaining charges. If
$\mathcal{Q}_{0}\neq 0$, we find the following condition

\begin{equation}
\label{eq:Q+Q-}
\mathcal{Q}_{+} \gtrsim \mathcal{Q}_{-}\, ,  
\,\,\,
\Rightarrow
\,\,\,
N_{\rm W} \gg N_{\rm F1}\, .
\end{equation}

All these conditions can be summarized into 

\begin{equation}
N_{\rm W} \gg N_{\rm F1} \gg N_{\rm S5}\, .  
\end{equation}

For the case $\mathcal{Q}_{0}=0$ the only conditions that need to be satisfied are
those affecting the moduli, namely $g_{s}\ll 1$, $k_{\infty}>1$. These
solutions, however, have many other problems: their metrics are singular at
$\rho=0$ in $d=10$, to start with and the reason why they are regular in $d=5$
is that the compactification radius is also singular there.

Finally, we must find if and when the solution of the first-order in $\alpha'$
equations that we have obtained can be considered a first-order in $\alpha'$
approximation to a solution of the full Heterotic Superstring effective
action. Clearly, this happens if and when the higher-order corrections to the
$\mathcal{Z}$-functions are very small, compared with the first-order
solution.

It is not easy to assess the relevance of higher-order corrections without
actually computing them, which becomes increasingly difficult. Since the
higher-order corrections of the action and equations of motion are expected to
contain powers of the first-order corrections, many of them codified in the
so-called ``$\hat{T}$-tensors'' and in the Chern-Simons terms present in
$\hat{H}$, it is reasonable to expect that the higher-order corrections will
be smaller than the first-order corrections if the first-order corrections are
small enough. Since the first-order corrections are proportional to the
$\hat{T}$-tensors and to the Chern-Simons terms, they will be small if the
later are also small. Actually, a necessary criterion for a supersymmetric
solution to be exact to all orders in $\alpha'$ is the vanishing of
$\hat{T}$-tensors and the Chern-Simons terms \cite{Bergshoeff:1992cw}. 

The origin of the $T$-tensors is the need to supersymmetrize the Yang-Mills
and Lorentz Chern-Simons terms. There may be other terms in the action with a
different origin such as the well-known $\zeta (3)R^{4}$ term, but very little
is known about them.  When $\mathcal{Q}_{0}\neq 0$, it is usually argued that
these terms as well as other invariants occur in the action as inverse powers
of $N_{\rm S5}$, once the factors of $\alpha'$ have been taken into
account. The consequence is that $\mathcal{Q}_{0}$ is usually taken to be
large so $N_{\rm S5}$ is very large.

However, we would like to stress that it is not enough to study the scalar
invariants constructed from the curvature or from the $T$-tensors because, as
discussed in the paragraph following Eqs.~(\ref{eq:T4})-(\ref{eq:T}), some
components of the curvature and of the $T$ tensors that occur in the equations
of motion and source the first-order $\alpha'$ corrections such as
$\hat{T}^{(2)}{}_{uu}$ disappear in the scalar invariants. Thus, even if all
the curvature invariants vanish, one can expect non-vanishing $\alpha'$
corrections to the solutions such as those occurring in $\mathcal{Z}_{+}$.

The 3 $\hat{T}$-tensors are defined in Eq.~(\ref{eq:Ttensors}) and their
values, computed for this kind of solutions in Ref.~\cite{Cano:2017qrq} to
$\mathcal{O}(\alpha'^{2})$, are given in Eqs.~(\ref{eq:T4})-(\ref{eq:T}). It
is convenient to analyze the corrections for the cases $\mathcal{Q}_{0}\neq 0$
and $\mathcal{Q}_{0} =0$ separately.

When $\mathcal{Q}_{0}\neq 0$, all the components of these tensors, except for
$\hat{T}^{(2)}{}_{uu}$, as well as the combined Yang-Mills- and
Lorentz-Chern-Simons terms, become arbitrarily small for $\kappa^{2}\sim
\mathcal{Q}_{0}$. In fact, in agreement with this, the correction to
$\mathcal{Z}_{0}$, which we write here for convenience

\begin{equation}
8\alpha' 
\left[
\frac{\rho^{2}+2\kappa^{2}}{(\rho^{2}+\kappa^{2})^{2}}
-
\frac{\rho^{2}+2\mathcal{Q}_{0}}{(\rho^{2}+\mathcal{Q}_{0})^{2}}
\right]\, ,
\end{equation}

\noindent
also becomes arbitrarily small in this limit, independently of the value of
$\mathcal{Q}_{0}$, which is usually necessary to assume very large.

For $\kappa^{2}=\mathcal{Q}_{0}$ all the components of the $T$-tensors, except
for $\hat{T}^{(2)}{}_{uu}$, and the correction of $\mathcal{Z}_{0}$ vanish
identically. As mentioned before, if we set
$\mathcal{Q}_{+}=\mathcal{Q}_{-}=0$ we recover the so-called ``symmetric
5-brane'' solution of Ref.~\cite{Callan:1991dj} which has been argued to be an
exact solution to all orders in $\alpha'$ ( $\hat{T}^{(2)}{}_{uu}=0$).  In the
general case we have to consider the effects of the non-vanishing
$\hat{T}^{(2)}{}_{uu}$. At first order, it sources the $uu$ component of the
Einstein equations, which only affect $\mathcal{Z}_{+}$. At higher order, it
cannot occur in any invariant, as we have explained. It can only appear
multiplied by invariants sourcing the same component of the Einstein
equations. Those invariants can be made as small as wanted with
$\kappa^{2}\sim \mathcal{Q}_{0}$ and, therefore, since the first-order
correction of $\mathcal{Z}_{+}$ ($f_{+}(\rho)$ in Eq.~(\ref{eq:f+2})) is
regular everywhere, it is reasonable to expect that the higher-order
corrections will also be finite but much smaller.

The conclusion, thus, is that for $\mathcal{Q}_{0}\neq 0$, and
$\mathcal{Q}_{0}\gg \alpha'$, taking $\kappa^{2}\sim \mathcal{Q}_{0}$ we get a
very good approximation to an exact solution of the Heterotic String effective
action.

For $\mathcal{Q}_{0}=0$ (small black holes) the corrections associated to the
gauge 5-brane are finite and at higher orders, multiplied by higher powers of
$\alpha'$, much smaller, but cannot be completely cancelled. We can simply
remove the gauge 5-brane to simplify the problem, eliminating these
corrections. The main problem, though, is the correction in $\mathcal{Z}_{+}$
associated to $\hat{T}^{(2)}{}_{uu}$, which diverges at the horizon and which
will diverge there at higher orders even if we multiply it by small numbers,
as long as they are non zero. The divergence of the first-order correction, by
itself, only indicates that the zeroth-order solution is not to be trusted at
the horizon. The first-order solution can be trusted if the rest of the
corrections vanish which, according to the previous discussions, may happen if
we remove the gauge 5-brane.

Nevertheless, as we have pointed out before, the small black-hole solutions
are singular in 10 dimensions and the $\alpha'$ correction to their entropy
seems to be divergent. Furthermore, their T-dual is singular because
$\mathcal{Q}_{-}'=-|\mathcal{Q}_{+}|<0$ and $\mathcal{Z}_{-}'$ with vanish at
$\rho^{2}=|\mathcal{Q}_{+}|$. These properties suggest that these solutions,
which may have negative mass in $d=5$, are not good solutions of Heterotic
String Theory.

\section{Discussion}
\label{sec-discussion}

In this paper we have computed explicitly the first $\alpha'$ corrections to a
3-charge 5-dimensional black hole to which we have added an $\mathrm{SU}(2)$
Yang-Mills instanton, and we have studied some of the effects that these
corrections have on the geometry, entropy and mass of the solutions. We have
also studied the effect of an $\alpha'$-corrected T-duality transformation in
the $\alpha'$-corrected solution, testing simultaneously the validity of our
solution and of the T-duality rules proposed, long time ago, in
Ref.~\cite{Bergshoeff:1995cg}. Studying the effect of these
$\alpha'$-corrected T-duality transformations requires the knowledge of
$\alpha'$-corrected solutions, which is very scarce in the literature.

The fact that the corrections can be computed explicitly is, by itself, a
remarkable fact. The computability of the corrections to the $\mathcal{Z}_{0}$
function is due to the surprisingly simple form of the Bianchi identity for the
configurations we have considered: a linear combination of Laplacians, a
``coincidence'' that can be generalized to more complicated supersymmetric
configurations \cite{Chimento:2018kop}.  

Finding the $\alpha'$ corrections to the S5-brane solution in presence of a
gauge 5-brane has also allowed us to gain better understanding of the
symmetric 5-brane solution found in Ref.~\cite{Callan:1991dj}.

Furthermore, we have shown how the $\alpha'$ corrections to the entropy of the
5-dimensional black holes can be computed using Wald's formula
directly in
10-dimensional language. Our calculation is very clean and transparent and
shows the relevance of the Lorentz-Chern-Simons term in the corrections and
the irrelevance of the curvature-squared terms (which was already known since
Ref.~\cite{Prester:2008iu}). Our results concerning the invariance under $\alpha'$-corrected T-duality 
(up to interchange of numbers of branes) of the family of solutions considered 
here, implies the invariance of the $\alpha'$-corrected entropy formula under the 
same transformations, in agreement with the results of Ref.~\cite{Edelstein:2018ewc}

Of course, we must compare our results with other results about higher-order
$\alpha'$ corrections to supersymmetric black-hole solutions in the
literature\footnote{See, for instance,
  Refs.~\cite{Mohaupt:2000mj,Prester:2010cw} and references therein.}. 

Most of the work done in this field deals with solutions to ungauged 4- and
5-dimensional $\mathcal{N}=2$ (8-supercharge) supergravities obtained via
Calabi-Yau compactifications from M-theory or type~II theories and (at least
some of) the 1st-order in $\alpha'$ corrections are said to be effectively
encoded in corrections to the prepotential (in $d=4$), for instance. This
obscures the origin of the corrections, which may or may not represent all
the corrections one finds in higher dimensions (see,
\textit{e.g.}~\cite{Hanaki:2006pj}), and makes it very difficult to say
anything about the relevance of corrections of orders higher than
1. Furthermore, the absence of non-Abelian fields forbids the use of the
``symmetric'' mechanism we have used to make very small or cancel many of the
corrections and argue the validity of our solution.\footnote{It has been
  explored with Abelian fields, though. See Ref.~\cite{Halmagyi:2016pqu} and
  references therein. An early use of this mechanism applied to a configuration
  related to that studied here can be found in Ref.~\cite{Kallosh:1994wy}, but
  it does not have enough charges to be a regular extremal black hole in lower
  dimensions. In a forthcoming publication we will show the relation between
  that configuration and the one studied here \cite{Chimento:2018kop}.} Finally, it
is unclear where the relevant contribution of 10-dimensional
Lorentz-Chern-Simons term to the first-order corrections is to be found in 4
or 5 dimensions. Thus, comparing our results with those obtained within this
approach is very difficult.

Some work has also been done using a 10-dimensional approach to the
computation of the corrections in Heterotic Superstring Theory,\footnote{See
  Ref.~\cite{Prester:2010cw} and references therein.} but only near-horizon
geometries were studied,\footnote{they have also been used in the context of
  the entropy-functional approach \cite{Sahoo:2006pm}.} while we have studied
and computed the corrections to the full black-hole geometry from infinity to
the horizon. While we have concluded that the parameters of interest are the
``near-horizon charges'', because they count the number of string-theory
objects sourcing the solution, of course, the total charges, measured at
infinity and these constants are related, and the relation can be computed
explicitly in our $\alpha'$-corrected solutions because they describe both
regions. Writing the entropy or the mass in terms of one or the other is a
matter of choice, but, after they are written in terms of numbers of branes
and other quantized quantities that are expected to be strictly positive, one
expects the solution to have sensible physical properties. It is not hard to
see that, when all the charges are different from zero, if the asymptotic
charges were identified with the quantized charges, it would be possible to find
negative-mass solutions.

The value found for the $\alpha'$-corrected entropy, Eq.~(\ref{eq:entropy})
seems to disagree with the value of the microscopic entropy computed in
Ref.~\cite{Castro:2008ys} as written in Ref.~\cite{Prester:2010cw}, but the
value of $\alpha'$ in that reference is 8 times ours and, therefore, they
coincide, although the route followed to arrive at the same result is totally
different.

The fact that the $\alpha'$ corrections associated to the torsionful spin
connection have the ``wrong sign'' as compared with those of the Yang-Mills
fields is clearly the source of some of this pathological behavior, already
hinted at by the results of Ref.~\cite{Hubeny:2004ji}, in which the
$\alpha'$-corrected black holes were shown to be repulsive. The addition of
Yang-Mills fields can correct some of these effects, making some of the
$\alpha'$ corrections very small or zero, but not all of them. It is, however,
likely, that a more general kind of Yang-Mills fields which give rise to
non-Abelian dyons in 5 dimensions can cancel all of them. Work in this
direction is in progress \cite{Chimento:2018kop}.

\section*{Acknowledgments}

The authors would like to thank S.~Chimento and A.~Ruip\'erez for many useful
conversations.  This work has been supported in part by the MINECO/FEDER, UE
grants FPA2015-66793-P and FPA2015-63667-P, by the Italian INFN and by the Spanish Research
Agency (Agencia Estatal de Investigaci\'on) through the grant IFT Centro de
Excelencia Severo Ochoa SEV-2016-0597.  The work of PAC was funded by
Fundaci\'on la Caixa through a ``la Caixa - Severo Ochoa'' international
pre-doctoral grant. PAC also thanks Perimeter Institute
``Visiting Graduate Fellows'' program. Research at Perimeter Institute is
supported by the Government of Canada through the Department of Innovation,
Science and Economic Development and by the Province of Ontario through the
Ministry of Research, Innovation and Science. TO wishes to thank
M.M.~Fern\'andez for her permanent support.

\appendix

\section{Connection, torsionful spin connection etc.}
\label{sec-connection}

In this appendix we are going to compute explicitly the connections and
curvatures of the ansatz Eq.~(\ref{10dmetric}). While that ansatz is
spherically symmetric in a 4-dimensional space, it is more convenient to do
some of the computations using a slightly more general ansatz and then
particularize to spherical symmetry. 

Thus, here, we are interested in 10-dimensional metrics of the form

\begin{equation}
ds^{2}
=
\frac{2}{\mathcal{Z}_{-}}du
\left[dv-\tfrac{1}{2}\mathcal{Z}_{+}du\right]
-\mathcal{Z}_{0} dx^{m}dx^{m} -dy^{i}dy^{i}\, ,
\end{equation}

\noindent
where $m,n,i,j=1,2,3,4$ and the functions
$\mathcal{Z}_{\pm},\mathcal{Z}_{0},H$ are functions on the first
4-dimensional space with coordinates $x^{m}$. Thus, the metric is independent
of the light-cone coordinates $u,v$ and of the 4 spatial coordinates $y^{i}$.

A simple choice of Zehnbein is 

\begin{equation}
\label{eq:zehnbein}
e^{+} = \mathcal{Z}_{-}^{-1}du\, , 
\quad 
e^{-} = dv -\tfrac{1}{2}\mathcal{Z}_{+}du\, , 
\quad 
e^{m} = \mathcal{Z}_{0}^{1/2}\, dx^{m}\, , 
\quad 
e^{i}=dy^{i}\, ,
\end{equation}

\noindent
and the inverse basis is 

\begin{equation}
e_{+} 
=
\mathcal{Z}_{-}(\partial_{u}+\tfrac{1}{2}\mathcal{Z}_{+}\partial_{v})\, ,
\quad 
e_{-} 
=
\partial_{v}\, ,
\quad
e_{m}
=
\mathcal{Z}_{0}^{-1/2}\partial_{m}\, ,
\quad
e_{i}=\partial_{i}\, ,
\end{equation}

\noindent
where $\partial_{m}\equiv \partial_{\underline{m}}$ and
$\partial_{i}\equiv \partial_{\underline{i}}$.

Using the structure equation $de^{a}=\omega^{a}{}_{b}\wedge e^{b}$ we find
that the non-vanishing components of the spin connection are given by 

\begin{equation}
\begin{array}{rclrcl}
\omega_{-+m} & = & \omega_{+-m} = \omega_{m+-} 
=
\tfrac{1}{2} \mathcal{Z}_{0}^{-1/2}
\partial_{m}\log{\mathcal{Z}_{-}}\, ,
\quad 
&
\omega_{+m+} 
& = & 
-\tfrac{1}{2}\mathcal{Z}_{-}\mathcal{Z}^{-1/2}_{0}
\partial_{m}\mathcal{Z}_{+}\, , 
\\
& & & & & \\
\omega_{mnp} 
& = &
\mathcal{Z}_{0}^{-3/2}\delta_{m[n}\partial_{p]} \mathcal{Z}_{0}\, .
& &  & 
\end{array}
\end{equation}

We are also interested in 3-form field strengths of the general form

\begin{equation}
\label{eq:generalH}
H 
=
du \wedge dv\wedge  d\mathcal{Z}^{-1}_{-}
+\star_{(4)}d\mathcal{Z}_{0}\, ,
\end{equation}

\noindent
where $\star_{(4)}$ is the Hodge dual in the first 4-dimensional space with
the orientation $\varepsilon^{\sharp 123}=+1$. Their non-vanishing flat components
are

\begin{equation}
\label{eq:Hcomponents}
H_{m+-}
=  
-
\mathcal{Z}_{0}^{-1/2}\partial_{m}\log{\mathcal{Z}_{-}}\, ,
\hspace{1cm}
H_{mnp}
=
\mathcal{Z}_{0}^{-1/2}\, 
\varepsilon_{mnpq}\partial_{q}\log{\mathcal{Z}_{0}}\, .
\end{equation}

Then, the non-vanishing flat components of torsionful spin connection
$\Omega_{(-)abc}\equiv \omega_{abc}-\tfrac{1}{2}H_{abc}$ are

\begin{equation}
\begin{array}{rclrcl}
\Omega_{(-)++m}
& = &
\tfrac{1}{2}
\mathcal{Z}_{-}\mathcal{Z}_{0}^{-1/2}\partial_{m}\mathcal{Z}_{+}\, ,
&
\Omega_{(-)+-m}
& = &
\mathcal{Z}_{0}^{-1/2}\partial_{m}\log{\mathcal{Z}_{-}}\, ,

\\    
& & & & & \\
\Omega_{(-)m+-}
& = & 
\Omega_{(-)+-m}\, ,
&
\Omega_{(-)mnp}
& = &
\mathcal{Z}_{0}^{-1/2}
(\mathbb{M}^{-}_{mq})_{np}\partial_{q}\log{\mathcal{Z}_{0}}\, ,
\end{array}
\end{equation}

\noindent
and those of the torsionful spin connection $\Omega_{(+)abc}\equiv
\omega_{abc}+\tfrac{1}{2}H_{abc}$ are given by

\begin{equation}
\begin{array}{rclrcl}
\Omega_{(+)++m}
& = &
\tfrac{1}{2}
\mathcal{Z}_{-}\mathcal{Z}_{0}^{-1/2}\partial_{m}\mathcal{Z}_{+}\, ,
&
\Omega_{(+)-+m}
& = &
\mathcal{Z}_{0}^{-1/2}\partial_{m}\log{\mathcal{Z}_{-}}\, ,
\\    
& & & & & \\
\Omega_{(+)mnp}
& = &
\mathcal{Z}_{0}^{-1/2}
(\mathbb{M}^{+}_{mq})_{np}\partial_{q}\log{\mathcal{Z}_{0}}\, ,
& & & \\
\end{array}
\end{equation}

\noindent
where the $4\times 4$ matrices $\mathbb{M}^{\pm}_{mq}$ are the self- and
anti-self-dual parts of the generators of SO$(4)$:

\begin{equation}
(\mathbb{M}_{mq})_{np} 
=
(\mathbb{M}_{np})_{mq} 
\equiv 
2\delta_{n[m}\delta_{q]p}\, ,
\hspace{1cm} 
\mathbb{M}^{\pm}_{mq}
\equiv 
\tfrac{1}{2}
\left(\mathbb{M}_{mq}\pm
\tfrac{1}{2}\varepsilon_{mqrs}\mathbb{M}_{rs}\right)\, .
\end{equation}

The components with curved indices are given by

\begin{equation}
\label{eq:torsionfulconnection}
\begin{array}{rclrcl}
\Omega_{(-)\underline{m}+-}
& = &
\partial_{m}\log{\mathcal{Z}_{-}}\, ,
&
\Omega_{(-)u-m}
& = &
-\mathcal{Z}_{0}^{-1/2}\partial_{m}\mathcal{Z}^{-1}_{-}\, ,
\\
& & & & & \\
\Omega_{(-)u+m}
& = &
\tfrac{1}{2}\mathcal{Z}_{0}^{-1/2}\partial_{m}\mathcal{Z}_{+}\, ,
\hspace{1cm}
&
\Omega_{(-)\underline{m}np}
& = &
(\mathbb{M}^{-}_{mq})_{np}\partial_{q}\log{\mathcal{Z}_{0}}\, ,
\end{array}
\end{equation}

\noindent
and

\begin{equation}
\begin{array}{rclrcl}
\Omega_{(+)u+m}
& = &
\tfrac{1}{2}
\mathcal{Z}_{0}^{-1/2}\partial_{m}\mathcal{Z}_{+}\, ,
&
\Omega_{(+)v+m}
& = &
\mathcal{Z}_{0}^{-1/2}\partial_{m}\log{\mathcal{Z}_{-}}\, ,
\\    
& & & & & \\
\Omega_{(+)mnp}
& = &
(\mathbb{M}^{+}_{mq})_{np}\partial_{q}\log{\mathcal{Z}_{0}}\, ,
& & & \\
\end{array}
\end{equation}

\subsection{Solving the Bianchi identity for $H$}
\label{sec-bianchiH}

Observe that $\Omega_{(-)\underline{m}np}$ coincides with the form of the
't~Hooft ansatz for $\mathrm{SU}(2)$ Yang-Mills multi-instanton solutions
using SO$(4)$ indices.\footnote{The same is true, with opposite self-duality,
  for $\Omega_{(+)\underline{m}np}$, but we will focus on
  $\Omega_{(-)\underline{m}np}$ only, because it is the one whose Chern-Simons
  3-form and curvature occur in the equations of motion.} Furthermore, this is
the only piece of $\Omega_{(-)\mu ab}$ that contributes to the
Lorentz-Chern-Simons term:

\begin{equation}
{\omega}^{{\rm L}}_{(-)}
= 
d{\Omega}_{(-)mn} \wedge 
{\Omega}_{(-)nm} 
+\tfrac{2}{3}
{\Omega}_{(-)mn} \wedge 
{\Omega}_{(-)np} \wedge
{\Omega}_{(-)pm}
=
\star_{(4)}d(\partial\log{\mathcal{Z}_{0}})^{2}
\, .
\end{equation}

\noindent
Then 

\begin{equation}
\label{eq:TrRR}
{R}_{(-)}{}^{{a}}{}_{{b}}\wedge 
{R}_{(-)}{}^{{b}}{}_{{a}}
=
d {\omega}^{{\rm L}}_{(-)}
= 
d\star_{(4)}d(\partial\log{\mathcal{Z}_{0}})^{2}
=
-\partial_{m}\partial_{m}(\partial\log{\mathcal{Z}_{0}})^{2}d^{4}x\, ,
\end{equation}

\noindent
where $d^{4}x$ is the volume form of $\mathbb{E}^{4}$. To obtain this expression we have used the local connection $\Omega_{(-)\underline{m}np}$ given in \eqref{eq:torsionfulconnection}, which is well defined in $\mathbb{R}^4$ except at the pole of $\mathcal{Z}_0$ at $\rho=0$, where it becomes singular. Since the quantity computed in \eqref{eq:TrRR} is gauge invariant, the result obtained is valid everywhere except at this isolated point, which is not covered by our local connection. Evaluating explicitly the right hand side, at zeroth-order in $\alpha'$, we obtain

\begin{equation}
-\partial_{m}\partial_{m}(\partial\log{\mathcal{Z}^{(0)}_{0}})^{2}
=
\partial_{m}\partial_{m} \left[
4\frac{\rho^{2}+2\mathcal{Q}_{0}}{(\rho^{2}+\mathcal{Q}_{0})^{2}}
-\frac{4}{\rho^{2}} \right]
=
\partial_{m}\partial_{m} \left[
4\frac{\rho^{2}+2\mathcal{Q}_{0}}{(\rho^{2}+\mathcal{Q}_{0})^{2}}
 \right]
-4 \delta^{(4)} (\rho) \, .
\end{equation}

\noindent
While the first term in that expression is a continuous, regular function, the second term just introduces a pointlike singularity at $\rho=0$ that, according to the preceding discussion, should be interpreted as spurious. 

Since the components of the 4-form ${R}_{(-)}{}^{{a}}{}_{{b}}\wedge {R}_{(-)}{}^{{b}}{}_{{a}}$ are continuous, at this stage it is clear that at this order in $\alpha'$ we have\footnote{This result is also obtained by performing a (singular) local Lorentz transformation that would render the torsionful spin connection regular at $\rho=0$, in virtue of the removable singularity theorem of Uhlenbeck Ref.~\cite{Uhlenbeck:1982zm}.}

\begin{equation}
{R}_{(-)}{}^{{a}}{}_{{b}}\wedge 
{R}_{(-)}{}^{{b}}{}_{{a}}
=
\partial_{m}\partial_{m} \left[
4\frac{\rho^{2}+2\mathcal{Q}_{0}}{(\rho^{2}+\mathcal{Q}_{0})^{2}}
 \right] dx^4
\equiv
-  \partial_{m}\partial_{m}  \left[ (\partial\log{\mathcal{Z}^{(0)}_{0}})^{2} \right]_{\backslash \odot} d^{4}x\, ,
\end{equation}

\noindent
where $\mathcal{Z}^{(0)}_{0}$ is the piece of
$\mathcal{Z}_{0}$ which is of zeroth order in $\alpha'$, which is the
harmonic function in $\mathbb{E}^{4}$ defined in Eq.~(\ref{eq:harmonicpieces}). Here we have introduced the symbols $\{ \backslash \odot \}$ to indicate that the (harmonic) singular term should be removed from the term within squared brackets.

It is convenient to use the 't~Hooft ansatz with SO$(4)$ indices for the gauge
field as well. We can write it in the form 

\begin{equation}
A=\mathbb{M}^{-}_{mp}\partial_{p}\log{\mathcal{Z}_{\rm YM}}dx^{m}\, ,  
\end{equation}

\noindent
where $\mathcal{Z}_{\rm YM}$ is the harmonic function on $\mathbb{E}^{4}$

\begin{equation}
\mathcal{Z}_{\rm YM} = 1+\frac{\kappa^{2}}{\rho^{2}}\, .
\end{equation}

\noindent
Using the result obtained for the $\omega^{L}_{(-)}$ \footnote{We have to take
  into account that the anti-self-dual SO$(4)$ generators have the
  normalization 
\begin{equation}
\mathrm{Tr}(\mathbb{M}^{-}_{mn}\mathbb{M}^{-}_{pq}) 
= 
-2(\mathbb{M}^{-}_{mn})_{pq}\, ,
\hspace{1cm}
[\mathbb{M}^{-}_{0i},\mathbb{M}^{-}_{0j}]
=
\varepsilon_{ijk}\mathbb{M}^{-}_{0k}\, ,
\hspace{.5cm}
i=1,2,3\, .
\end{equation}
Therefore, using the above representation, 
%
\begin{eqnarray}
F
& = &
dA+A\wedge A\, ,
\\
& & \nonumber \\
\omega^{{\rm YM}}
& = & 
-\mathrm{Tr}\left[dA\wedge {A}
+\tfrac{2}{3}{A}\wedge{A}\wedge{A}
\right]\, ,
\\
& & \nonumber \\
F^{A}\wedge F^{A}
& = &
-\mathrm{Tr} F\wedge F\, .    
\end{eqnarray}
}

\begin{eqnarray}
\omega^{{\rm YM}}
& = & 
-\star_{(4)}d(\partial\log{\mathcal{Z}_{\rm YM}})^{2}
\, ,
\\
& & \nonumber \\
F^{A}\wedge F^{A}
& = & 
d \omega^{{\rm YM}}
=
 \partial_{m}\partial_{m} \left[(\partial\log{\mathcal{Z}_{\rm YM}})^{2}  \right]_{\backslash \odot} d^{4}x\, ,
\end{eqnarray}

\noindent
where, following the same reasoning as before, the singular contribution must be removed. Thus, taking into account the general form of the 3-form $H$ in Eq.~(\ref{eq:generalH}), the Bianchi identity of the 3-form field strength
Eq.~(\ref{eq:BianchiH}) can be written in the form 

\begin{equation}
\label{eq:BIanchidigested1}
-\partial_{m}\partial_{m} 
\left\{
\mathcal{Z}_{0}
+2\alpha'  
\left[(\partial\log{\mathcal{Z}_{\rm YM}})^{2}
-(\partial\log{\mathcal{Z}^{(0)}_{0}})^{2}
\right]_{\backslash \odot}
\right\}
=
0\, ,  
\end{equation}

The above equation is solved by

\begin{equation}
\mathcal{Z}_{0}
=
\mathcal{Z}^{(0)}_{0} 
+2\alpha'
\left[(\partial\log{\mathcal{Z}^{(0)}_{0}})^{2} 
-(\partial\log{\mathcal{Z}_{\rm YM}})^{2}
\right]_{\backslash \odot}
+\mathcal{O}(\alpha'^{2})\, ,
\end{equation}

\noindent
where we have used that $\mathcal{Z}_{0} =
\mathcal{Z}^{(0)}_{0}+\mathcal{O}(\alpha')$. In the language of Section~\ref{sec-corrections},

\begin{equation}
f_{0}(\rho)
=
2
\left[(\partial\log{\mathcal{Z}^{(0)}_{0}})^{2} 
-(\partial\log{\mathcal{Z}_{\rm YM}})^{2}
\right]_{\backslash \odot}
=
8\left[\frac{\rho^{2}+2\kappa^{2}}{(\rho^{2}+\kappa^{2})^{2}}
-\frac{\rho^{2}+2\mathcal{Q}_{0}}{(\rho^{2}+\mathcal{Q}_{0})^{2}}
\right]\, ,
\end{equation}

\noindent
which is the same result as in Eq.~(\ref{eq:f0solution}). Upon substitution in the Bianchi identity, it reduces to the Laplacian of a
harmonic function on $\mathbb{E}^{4}$:

\begin{equation}
\label{eq:BIanchidigested2}
-\partial_{m}\partial_{m} \mathcal{Z}_{0}=0\, .
\end{equation}

\noindent
As usual, this equation is not satisfied at the singularities of the harmonic
function and the corresponding $\delta$-functions will give contributions to
the S5-brane charge (see Eq.~(\ref{eq:S5branecharge})).

Using these results in the definition of $H$ Eq.~(\ref{Hdef}) we arrive at the
following equation for the Kalb-Ramond 2-form $B$:

\begin{equation}
  dB= d\left[\mathcal{Z}_{-}^{-1}du\wedge dv\right]
  +\star_{(4)}d\mathcal{Z}^{(0)}_{0}\, .  
\end{equation}

\noindent
The integrability condition is satisfied if $\mathcal{Z}^{(0)}_{0}$ is
harmonic in $\mathbb{E}^{4}$ and for the value in
Eq.~(\ref{eq:harmonicpieces}), it is given by

\begin{equation}
\label{eq:B}
B
= 
\mathcal{Z}^{-1}du\wedge dv
+\tfrac{1}{4}\mathcal{Q}_{0}\cos{\theta}d\varphi\wedge d\psi\, ,  
\end{equation}

\noindent
and receives no $\alpha'$-corrections to this order.


\end{document}